%% file: bare_jrnl.tex
\documentclass[journal]{IEEEtran}
\usepackage{amssymb}
\usepackage{xspace}
\usepackage{color}
\usepackage{graphicx}
\usepackage{amsmath} 
\usepackage{cleveref}
\usepackage{todonotes}
\usepackage{url}
\usepackage{dsfont}
\usepackage{multirow}
\usepackage{subfigure}
\usepackage{colortbl}

\def\oursystem{LensDroid\xspace}
\def\ie{\textit{i.e.}\xspace}
\def\etc{\textit{etc.}\xspace}
\def\eg{\textit{e.g.}\xspace}

\usepackage[ruled]{algorithm}
\usepackage[noend]{algpseudocode}
\renewcommand{\algorithmicrequire}{\textbf{Input:}}  
\renewcommand{\algorithmicensure}{\textbf{Output:}} 

\begin{document}

\title{Detecting Android Malware by Visualizing App Behaviors from Multiple Complementary Views}

\author{Zhaoyi Meng,
        Jiale Zhang,
        Jiaqi Guo,
        Wansen Wang,
        Wenchao Huang,
        Jie Cui,
        Hong Zhong,
        and Yan Xiong
\thanks{Zhaoyi Meng, Jiale Zhang, Wansen Wang, Jie Cui, Hong Zhong are with the School of Computer Science and Technology, Anhui University, Hefei, 230039, China. Wenchao Huang, Jiaqi Guo, Yan Xiong are with the School of Computer Science and Technology, University of Science and Technology of China, Hefei 230027, China. Corresponding authors:
Wansen Wang, Wenchao Huang. (e-mail: 
zymeng@ahu.edu.cn; 
23762@ahu.edu.cn; huangwc@ustc.edu.cn
)}}

\maketitle

\input{abstract}
\input{introduction}
\input{motivation}

\input{design}

\input{evaluations}
\input{discussion}
\input{relatedwork}
\input{conclusion}

\bibliographystyle{IEEEtran}
\bibliography{IEEEabrv, mybibfile}

\end{document}

%% file: abstract.tex

\begin{abstract}

Deep learning has emerged as a promising technology for achieving Android malware detection.
To further unleash its detection potentials, software visualization can be integrated for analyzing the details of app behaviors clearly.
However, facing increasingly sophisticated malware, existing visualization-based methods, analyzing from one or randomly-selected few views, can only detect limited attack types.
We propose and implement \textit{\oursystem}, a novel technique that detects Android malware by visualizing app behaviors from multiple complementary views.
Our goal is to harness the power of combining deep learning and software visualization to automatically capture and aggregate high-level features that are not inherently linked, thereby revealing hidden maliciousness of Android app behaviors.
To thoroughly comprehend the details of apps, we visualize app behaviors from three related but distinct views of behavioral sensitivities, operational contexts and supported environments.
We then extract high-order semantics based on the views accordingly.
To exploit semantic complementarity of the views, we design a deep neural network based model for fusing the visualized features from local to global based on their contributions to downstream tasks.
A comprehensive comparison with five baseline techniques is performed on datasets of more than 51K apps in three real-world typical scenarios, including overall threats, app evolution and zero-day malware.
The experimental results show that the overall performance of \oursystem is better than the baseline techniques.
We also validate the complementarity of the views and demonstrate that the multi-view fusion in \oursystem enhances Android malware detection.

\end{abstract}

%% file: introduction.tex

\section{Introduction}

\IEEEPARstart{m}{alware} attacks pose a serious threat to the development of the Android app ecosystem.
AV-TEST institute reports 35,362,643 malware and 26,944,746 potentially unwanted apps under Android between January and October 2024~\cite{AV-TEST2023}.
The apps can perform privacy theft, bank/financial stealing, privilege escalation, \textit{etc.}, which affect the daily work and life of Android users.
With the evolution of Android malware, attackers tend to hide their real intentions in various types of files of APKs.
For example, in May 2023, Trend Micro reported that Guerrilla malware is preinfected on 8.9 million Android devices~\cite{TrendReports}.
Through further research, it was found that the malware can use \textit{libandroid\_runtime.so} file to decrypt and execute \textit{.dex} files that contain malicious behaviors.

Manually crafting features is one of the most common ways for conventional signature-based and machine learning(ML)-based schemes to build Android malware detectors.
Signature-based methods find specific patterns in app bytecode~\cite{shi2020vahunt,arora2019permpair} or runtime behaviors~\cite{alhanahnah2020dina,tsutano2019jitana}.
However, the discovery of the patterns heavily depends on the researchers' experience,
so each of the methods can only identify a limited type of Android attacks. Traditional ML-based approaches extract features from apps (\textit{e.g.}, API usages~\cite{zhang2020enhancing,samhi2022difuzer}, behavioral relations~\cite{hou2017hindroid,onwuzurike2019mamadroid}, contents~\cite{arp2014drebin}), and apply standard ML algorithms (\textit{e.g.}, support vector machine) to train classifiers.
Nevertheless, selecting the hand-crafted features is time-consuming and labor-intensive.
Furthermore, the features used in each work capture only partial semantics of app behaviors.
Moreover, running all possible detection techniques for each app is costly in terms of both time and computational resources.

Inspired by the momentum of deep learning (DL) in multifarious areas, many schemes leverage the cost-effective technique to obtain hidden features automatically from Android apps.
Software visualization, by clearly revealing details of app behaviors, empowers the DL technique to further unleash its potentials in extracting semantic and structural features that are not inherently linked~\cite{chen2020software}.
Specifically, some works convert different types of app code into images~\cite{sun2021android,he2023resnext+}, graph representations~\cite{he2022msdroid,liu2023enhancing} or opcode-related structures~\cite{darem2021visualization,jeon2020malware}, and then train DL models (\textit{e.g.}, multi-layer perceptron (MLP), convolutional neural network (CNN), graph convolutional network (GCN)) to capture effective features for achieving malware detection.
However, with the continuous proliferation and rapid evolution of Android malware, it is difficult to detect various attack behaviors by relying on one (or randomly-selected few) of visualization-based methods.

To solve the problem above, it is promising to visualize and assess an app from multiple representative views.
Specifically, each visualization-based approach analyzes an app from a given view, which is capable of profiling certain characteristics of the app and meanwhile has its limitations.
For example, image visualizations for \textit{.so} files are amenable to detecting unwanted intentions within native code, but are incapable of identifying Java-based attacks. 
Therefore, a proper integration of multiple valuable views would help integrate strengths of the incorporated visualization-based approaches and exploit their complementary nature for disclosing hidden maliciousness within apps precisely and effectively.

\begin{table*}[t]
\caption{Statistics on Our Randomly Collected Datasets.}
\label{tab:dataset}
\centering
\resizebox{\linewidth}{!}{
\begin{tabular}{|c|c|c|c|c|c|c|c|c|c|}
\hline
\multirow{2}{*}{\textbf{Label}} & \multirow{2}{*}{\textbf{Year Range}}  & \multirow{2}{*}{\textbf{Source}} &   \multirow{2}{*}{\textbf{\# Apps}}   & \multicolumn{2}{c|}{\textbf{Callgraph}}       & \multirow{2}{*}{\textbf{\begin{tabular}[c]{@{}l@{}}Avg Size of\\ Opcode (KB)\end{tabular}}} & \multicolumn{3}{c|}{\textbf{Avg Size of Artifacts (KB)}}                                                                       \\ \cline{5-6} \cline{8-10} 
                                &          &                             &                               & \multicolumn{1}{l|}{\textbf{Avg \# Nodes}} & \textbf{Avg \# Edges}  &                                                                                        & \multicolumn{1}{c|}{\textbf{.dex}}            & \multicolumn{1}{c|}{\textbf{.xml}}            & \textbf{.so}              \\ \hline
\multirow{2}{*}{Malicious}      & 2010-2012                                                         & \begin{tabular}[c]{@{}l@{}}Drebin \&\\ AndroZoo\end{tabular}  & 8622  & 2314    & 18647       & 49.07        & 618.39   & 96.05    & 1984.89   \\ \cline{2-10} 
                                & 2018-2022                                                         &  AndroZoo  & 16810   & 1997 & 13069 & 73.39       & 6819.05 & 1098.48 & 15220.55 \\ \hline
\multirow{2}{*}{Benign}       & 2010-2012                                                    &  AndroZoo & 4893  & 2446 & 18531 &  71.82   &  984.84 &  188.53 & 5060.89 \\ \cline{2-10}
& 2018-2022                                                        &  AndroZoo & 20840 & 6522 & 58896 & 494.54    & 6658.09 & 421.20  & 18024.22 \\ \hline
\end{tabular}
}
\end{table*}

We propose and implement \textit{\oursystem}, a novel technique that detects Android malware by visualizing app behaviors from multiple complementary views.
Our goal is to harness the power of combining DL and software visualization to automatically capture and aggregate high-level features that are not inherently linked, thereby revealing hidden maliciousness of Android app behaviors.
To the best of our knowledge, \oursystem is the first to combine deep multi-view learning with software visualization for Android malware detection.

To understand the details of apps comprehensively, we decouple intricate factors that exist during the analysis of app behaviors into three related but distinct views, including behavioral sensitivities, operational contexts and supported environments.
Specifically, behavioral sensitivities highlight essential intentions of app behaviors, operational contexts enrich detailed associations around the key points that make up the intentions, and supported environments provide informative supplements for runtime execution of the behaviors.
Semantic information respectively obtained from the three views contributes to revealing maliciousness hidden in app code at different levels.
The appropriate fusion of the information can further promote complementarity among the views, and hence benefit the performance of fine-grained and precise assessments of Android app behaviors.

To implement \oursystem based on the decoupled views, we visualize apps to acquire three features correspondingly: (1) an abstract API callgraph for depicting the sensitivity in app code at the function level, (2) an opcode-gram-based matrix for aggregating the composition of opcode subsequences with a given length at the instruction level, (3) a binary-transformed image for presenting content layouts of multi-format artifacts at the bit level.
To extract high-order semantics from the heterogeneous features, we make vectorial representations separately by selecting suitable DL techniques according to their structural and content characteristics.
To exploit semantic complementarity of the views, we design an end-to-end model, using a deep neural network (DNN) combined with the multi-modal factorized bilinear (MFB) pooling and the multi-head self-attention mechanism, to fuse the vectors from local to global based on their contributions to app classification tasks.

Our main contributions are summarized as follows:
\begin{itemize}
\item We propose and implement \oursystem, a novel technique that precisely detects Android malware by visualizing app behaviors from multiple complementary views, and automatically capturing and aggregating high-level features that are not inherently linked.
\item We dissect app behaviors from three related but distinct views of behavioral sensitivities, operational contexts and supported environments to enhance malware detection.
\item Our comprehensive evaluations on more than 51K real-world apps with five baselines demonstrate the effectiveness of \oursystem and the complementarity of the views.
\end{itemize}

The paper is structured as follows: Section II motivates our work.
Section III details the methodology.
Section IV reports our experimental results.
Section V discusses our limitations.
Section VI shows the related work, and Section VII concludes.

%% file: motivation.tex

\section{Motivation}\label{sec:motivation}

To motivate our work, we make fine-grained statistics from the aspects of callgraphs, opcode and artifacts respectively on large-scale apps and then explain why the selected three views help identify Android malware.

As listed in \Cref{tab:dataset}, we collect 51165 samples belonging to multiple years from AndroZoo~\cite{allix2016androzoo} and Drebin~\cite{arp2014drebin}.
Specifically, we randomly gather 45875 real-world apps including 25733 benignware samples and 20142 malware samples from AndroZoo.
To enrich our datasets, we also obtain 5290 malware samples from a typcial dataset named Drebin.
We get the listed statistics based on off-the-shelf tools~\cite{DBLP:conf/pldi/ArztRFBBKTOM14,Androguard,APKTool}.
The details for handling the datasets are depicted in \Cref{subsubsec:dataset}.

We have the following two findings from the statistics:
\begin{itemize}
    \item The average number of nodes and edges in callgraphs, as well as the average size of opcode produced from benignware are much larger than those from malware between 2018 and 2022, whereas these differences are not evident from 2010 to 2012.
    For example, between 2018 and 2022, the callgraphs of benignware samples contain 58896 edges in average, which are about 450\% of the average number of edges for malware samples.
    In comparison, between 2010 and 2012, the average number of edges in the callgraphs of benignware samples is similar to those in malware samples.
    This suggests that the differences in callgraphs and opcode between benignware and malware are growing over time, and thereby can serve as important clues for detection efforts. 
    
    \item The average size of artifacts (\ie, the files of \textit{.dex}, \textit{.so}, \textit{.xml}) grows larger over time, especially for malware.
    For example, the average size of \textit{.so} files in the malware samples generated between 2018 and 2022 are about 767\% larger than those in the malware samples generated between 2010 and 2012.
    Similarly, the average size of \textit{.xml} and \textit{.dex} files also increases greatly.
    It implies that recent malware developers are likely to hide more attack intentions into multiple artifacts.
    Therefore, we also need to focus on the artifacts in malware detection.
\end{itemize}

Based on the findings above, we notice that it is essential to scrutinize apps using callgraphs, opcodes, and artifacts simultaneously.
The information from the sources contains insightful semantics of app behaviors at interrelated levels, so the appropriate fusion of them helps achieve a comprehensive malware detection.
Software visualization can be utilized on the views to clarify hidden maliciousness of apps more clearly.

%% file: design.tex
\begin{figure}[tb] 
\centering\includegraphics[width=\linewidth]{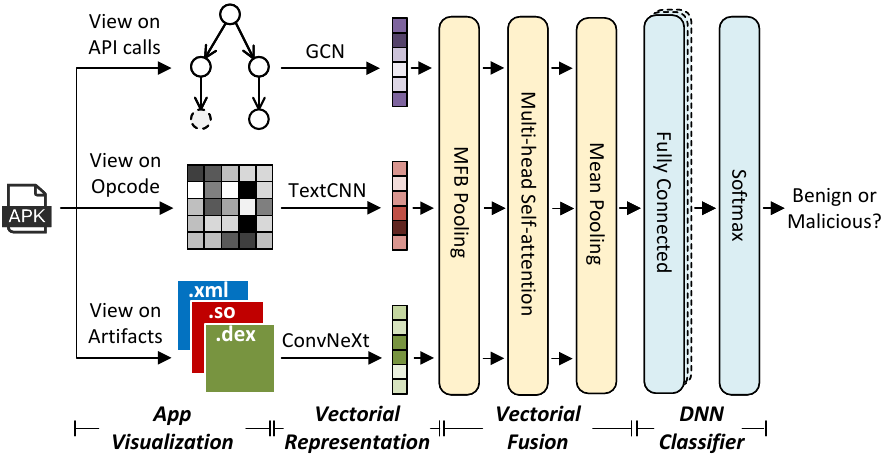}
\caption{The overall architecture of \oursystem.}
\label{overview11}
\end{figure}

\section{Design of \oursystem}
\subsection{Overview}\label{overview}
\Cref{overview11} depicts the overall architecture of \oursystem, which has four modules as follows.
\begin{itemize}
    \item \textbf{App Visualization:} To depict details of app behaviors comprehensively, the module generates visualized results (\ie, abstract API callgraphs, opcode-gram-based matrices, and binary-transformed images) from three interrelated views (\ie, behavioral sensitivities, operational contexts and supported environment) accordingly.
    The results work well with DL techniques for mining hidden semantics within app code in the following.
    \item \textbf{Vectorial Representation:} To learn high-order semantics under each view deeply, the module performs an appropriate DL methods (\eg, GCN~\cite{kipf2016semi}, TextCNN~\cite{kim2014convolutional}) for each visualized result based on its feature extraction requirements and then generates corresponding vectors.
    The vectors carrying rich information from the complementary views are fed to the next module separately.
    \item \textbf{Vectorial Fusion:} To exploit the complementary nature of different views for disclosing hidden maliciousness within app code, the module leverages a stepwise strategy with MFB pooling~\cite{yu2017multi} and multi-head self-attention mechanism~\cite{vaswani2017attention} to fuse the feature vectors from local to global.
    The fused vectors profiling app behaviors precisely are used in the following app classification.
    \item \textbf{DNN Classifier:} The module trains a standard DNN model to classify benignware and malware. As demonstrated in the proposed works~\cite{kim2018multimodal,qiu2022cyber}, DNN is typical and effective to deal with high-dimensional data about Android malware classification.
    Our classifier outputs the classification results with the Softmax function.
\end{itemize}

\subsection{Multi-view Feature Extraction and Representation}

\subsubsection{Feature of Abstract API Callgraphs}
We capture an app’s behavioral sensitivities using a structural API callgraph.
Specifically, many factors relate to behavioral sensitivities, where APIs are the most important parts because they enable direct access to critical functionalities and resources of operating systems~\cite{hou2017hindroid}.
Moreover, an individual API, whether sensitive or not, has limited functionality, while sophisticated behaviors are implemented from the combined use of multiple APIs~\cite{he2022msdroid}.
Thus, we use the callgraph to obtain semantic and structural relations among the calls of various APIs.

We extract the callgraph by off-the-shelf interfaces of FlowDroid~\cite{DBLP:conf/pldi/ArztRFBBKTOM14}.
The graph expresses call relations among different types of APIs, including sensitive APIs and normal APIs, in app code. 
However, it is imprecise to measure similarities of behavioral sensitivities between apps based on the callgraph directly.
With the development of Android platforms, many APIs are proposed or updated, and meanwhile, some APIs are deprecated.
Considering that apps are developed by different developers or at a different time, the APIs used to implement similar functionalities may be discrepant.

\begin{algorithm}[t]
  \caption{The process of API abstractions}
  \label{alg:APIEncoding}
  \small
   \begin{algorithmic}[1]
        \State{\algorithmicrequire{ an API set \textbf{A}, a mapping \textbf{M} between APIs and permissions, a mapping \textbf{P} between permissions and protection levels}}
        \State{\algorithmicensure{ the abstraction results \textbf{R} for the APIs in \textbf{A}}}
        \State{\textbf{G} = $\emptyset$}
        \For{each (m$_{api}$, m$_{per}$) $\in$ \textbf{M}}
            \State{(\textit{subfamily}, \textit{APIName}) $\leftarrow$ \textit{getInfo(}m$_{api}$\textit{)}}
            \State{\textit{subsignature} $\leftarrow$ \textit{infoCombination(}\textit{subfamily}, \textit{APIName}\textit{)}}
            \State{\textit{protectionLevels} $\leftarrow$ \textit{lookUp(}\textbf{P}, m$_{per}$\textit{)}}
            \State{// Using one-hot encoding based on the existence of the }
            \Statex{\qquad attributes in \textit{protectionLevels}}
            \State{\textit{vec} $\leftarrow$ \textit{protectionLevelEncoding(protectionLevels)}}
            \State{\textbf{G} $\leftarrow$ \textbf{G} $\cup$ (\textit{subsignature}, \textit{vec})}
        \EndFor

        \For{each \textit{api} $\in$ \textbf{A}}
            \State{(\textit{subfamily'}, \textit{APIName'}) $\leftarrow$ \textit{getInfo(}\textit{api)}}
            \State{\textit{subsignature'} $\leftarrow$ \textit{infoCombination(}\textit{subfamily'}, \textit{APIName'}\textit{)}}
            \State{\textit{vec'} $\leftarrow$ \textit{lookUp(}\textbf{G}, \textit{subsignature'}\textit{)}}
            \State{\textbf{R} $\leftarrow$ \textbf{R} $\cup$ (\textit{api}, \textit{vec'})}
        \EndFor
   \end{algorithmic}
\end{algorithm}

\begin{figure*}[tb]
\centering\includegraphics[width=\linewidth]{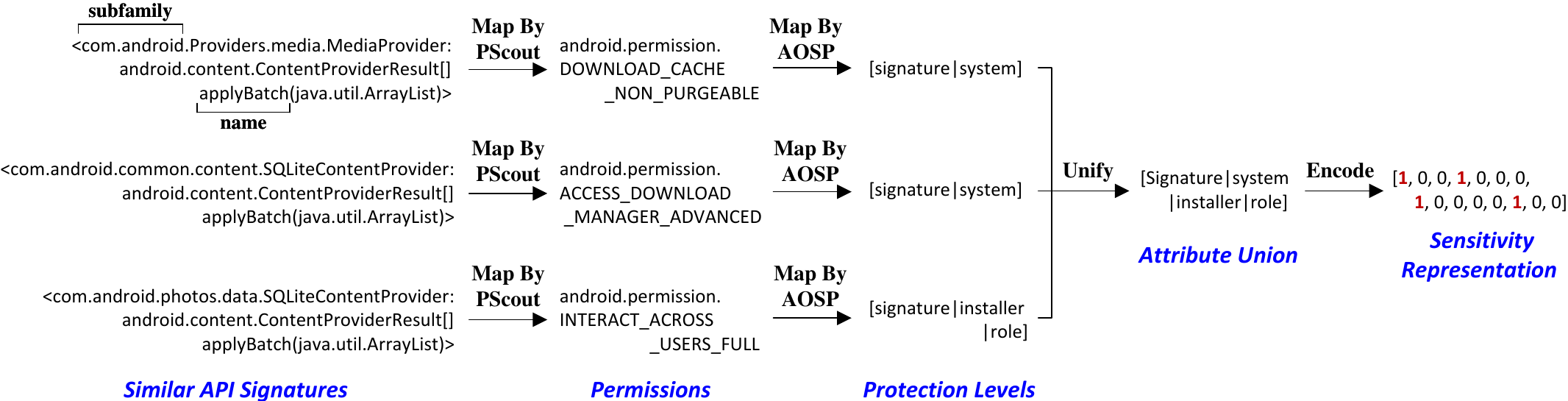}
\caption{An example for abstracting three similar APIs.}
\label{Example4APIEncoding}
\end{figure*}

To provide resilience to API changes and capture semantic commonalities between different callgraphs~\cite{onwuzurike2019mamadroid}, as depicted in \Cref{alg:APIEncoding}, we abstract the API calls in callgraphs into their sensitivity representations, and then generate abstract API callgraphs.
Specifically, we first obtain the mappings \textbf{M} and \textbf{P} from PScout~\cite{au2012pscout} and Android Open Source Project (AOSP) respectively.
Then we iterate through each element in \textbf{M} to extract a subsignature of each API and the corresponding protection levels based on \textbf{P} (Lines 5-8).
The subsignature is a combination of the first two fields of the class name (\eg, \textit{android.telephony}) and the method name (\eg, \textit{getDeviceId}) of an API.
The subsignature helps cluster the APIs for similar operations on similar objects, and is added and removed less frequently than single API calls.
Protection levels characterize the potential risk implied in each Android permission with attributes (\eg, \textit{signature$|$dangerous}).
If a subsignature corresponds to multiple groups of protection levels, we take the union set on the included attributes to integrate sensitive semantics of similar operations (Lines 9-10).
We next get protection levels for the APIs \textbf{A} in callgraphs, and vectorize them with one-hot encoding (Lines 11-14).
Note that we set the vector's length as 15 because the apps in our datasets cover 15 attributes of protection levels.
Finally, we acquire the sensitivity representations \textbf{R} for each API in \textbf{A} (Line 15).

~\Cref{Example4APIEncoding} exemplifies our abstractions on three similar APIs, which are used to batch process the data from media files, photos and SQLite databases respectively.
To identify the hidden maliciousness of the three APIs, the detailed object that each of the APIs operates on is not critical relatively. 
Instead, we should be more concerned about semantic sensitivity of this kind of APIs.
Hence, we group the three APIs because they all have the same subsignature (\ie, \textit{com.android \&\& applyBatch}).
To get the representation of their sensitivity, we unify protection levels of the APIs by off-the-shelf mappings from PScout and AOSP, and then perform the one-hot encoding.
It can be known by this example that every element in the sensitivity representation of a normal API is 0.

After finishing the API abstraction, the abstract API callgraph is constructed.
We then apply GCN~\cite{kipf2016semi}, one of the most useful techniques for graph node embedding, to collect neighbor information about call relations and sensitivities between APIs iteratively.
To obtain the embedded result for the whole callgraph, we aggregate the vectors of the nodes by calculating the average value element-wise.

\begin{figure}[tb]
\centering\includegraphics[width=0.9\linewidth]{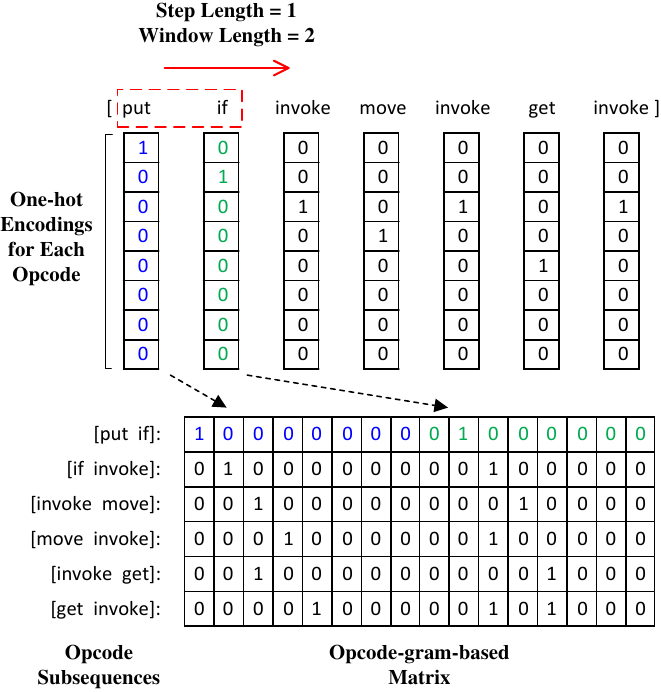}\label{fig:len}
\caption{A conversion from an opcode sequence to an opcode-gram-based matrix where the length of the sliding window is 2 and the step length is 1.}
\label{opcode}
\end{figure}

\subsubsection{Feature of Opcode-gram-based Matrices}\label{sec:context}
We obtain operational contexts of app behaviors from \textit{smali} code of apps.
The \textit{smali} code is the intermediate but interpreted code between Java and Dalvik Virtual Machine~\cite{hou2017hindroid}.
Each \textit{smali} instruction in \textit{smali} code consists of an opcode and multiple operands.
Based on that, we leverage APKTool~\cite{APKTool} to disassemble the \textit{.dex} file(s) of an app as a set of \textit{smali} code, where each Java class in the file(s) is translated to a \textit{smali} file.

To eliminate noise and improve effectiveness in the following malware detection, we first excerpt the content involving method bodies in each \textit{smali} file by keyword pairs \texttt{.method} and \texttt{.end method}.
The excerption is to remove the instructions about variable declarations, file sources, \etc
We then only remain opcode while discard operands for each \textit{smali} instruction.
The discarding is to remove annotations, the detailed values in instructions, \etc
All the removed contents above does not describe key contextual semantics of app behaviors directly.
In comparison, the preserved opcode complements semantic information around each API.
In other words, the sequence of opcode can serve as a bridge to explain the executing processes between successive APIs.

To extract semantic features of operational contexts, we vectorize the opcode.
Specifically, we alphabetize \textit{smali} files of an app and gather an opcode sequence line by line from each file.
The elements in the sequence are categorized into eight sets, including \textit{Move}, \textit{Get}, \textit{Put}, \textit{If}, \textit{Goto}, \textit{Invoke}, \textit{Return} and \textit{Separator}.
The first seven sets are the instruction types of opcode~\cite{qiu2022cyber}, and the last set is to separate opcode subsequences from different method bodies.
Therefore, we encode each of the elements in a one-hot vector of length 8.

Contextual semantics of APIs involve to the organized functionalities of multiple operations, so we convert the opcode sequence to an opcode-gram-based matrix.
This matrix encapsulates information from consecutive opcode elements.
Specifically, we leverage a sliding window to obtain successive and overlapping opcode subsequences horizontally, and then combine representation vectors of the subsequences vertically.
As exemplified in~\Cref{opcode}, the upper part depicts an opcode sequence with seven elements and their one-hot vectors.
The red dotted box is the sliding window for the subsequence extraction.
The length of the window is decided experimentally in \Cref{evaluation}.
As the window moves right, the subsequences are obtained and shown in the lower left.
We concatenate the one-hot vectors of the elements in each subsequence, and arrange the concatenations to form the matrix.

We adopt TextCNN~\cite{kim2014convolutional} to capture local correlations of the opcode subsequences in the matrix, \ie, contexts of sequential opcode, and generate the representative vector for an app.

\begin{figure}[tb]
\centering\includegraphics[width=0.7\linewidth]{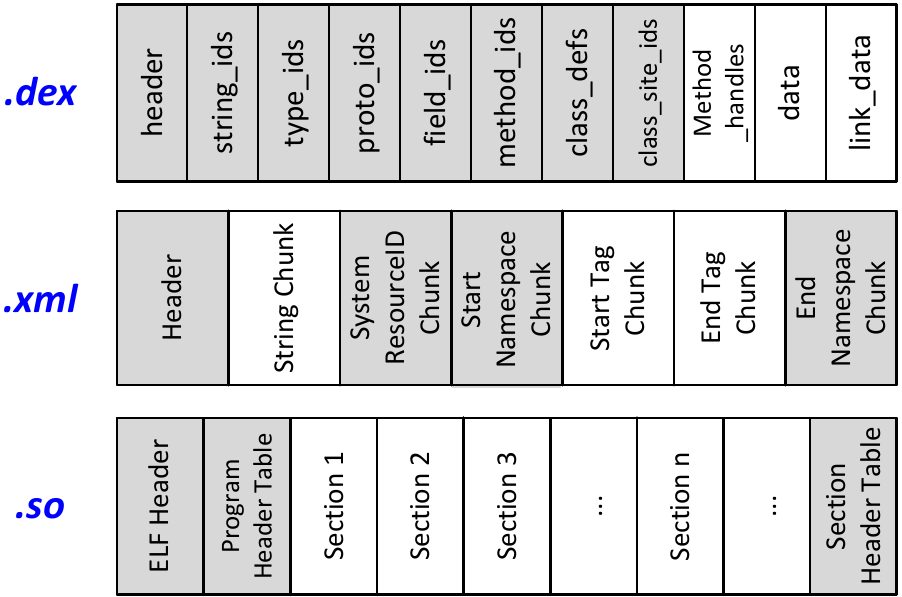}
\caption{The structures of three types of artifacts, where the white sections are preserved and the gray sections are regarded as noise. }
\label{dex}
\end{figure}

\subsubsection{Feature of Binary-transformed Images}
We acquire supported environments for app behaviors by combining three artifacts, \ie, \textit{.dex}, \textit{.xml} and \textit{.so} binary files, in an APK.
The three types of files imply different information about bytecode, configurations and native code of an app respectively. 
In previous works, some useful evidence is lost when certain files are not considered~\cite{samhi2022difuzer,he2022msdroid,2017DroidNative}, which may reduce the effectiveness of malware detections~\cite{sun2021android}.
It is still challenging to unify the three types of information directly.
Therefore, we make binary-transformed images from the three Android artifacts and learn the nature of app behaviors with DL techniques.
That is capable of aggregating semantic features that are hard to model and extract explicitly, instead of consuming heavyweight computations to make fine-grained analysis (\eg, data-flow paths, control-flow dependencies).

To enhance the accuracy of image representations for each app’s artifacts, we first preprocess the file contents by removing noise, as indicated by the grayed areas in \Cref{dex}. This ensures that only data directly related to app behaviors is retained.
For example, the structure of the \textit{.dex} file is divided into three areas: file header, index area and data area.
The contents in the file header and the index area do not portray app behaviors directly, so we regard them as noise that may lower the effectiveness of malware detection.
Similarly, the \textit{.xml} file and the \textit{.so} file also consist of data areas and metadata areas as well.
We therefore remove the noisy contents by parsing the file structures and then locating the target areas.

After that, we aggregate informative semantics within the three types of artifacts as follows.
Motivated by the excellent ability of DL techniques to capture detailed information at different scales in image recognition tasks, we generate a RGB image for each app where each color channel corresponds to a type of artifact.
Specifically, for each artifact, we first read its contents sequentially as a bitstream and divide them into consecutive 8-bit subsequences.
Afterwards, each subsequence is converted into a decimal number ranging from 0 to 255.
Based on the process, an artifact is converted into a 1-dimensional vector of decimal numbers.
We next resize the one-dimensional vector to a 2-dimensional matrix with a fixed width.
The matrix is treated as an image in a color channel.

When a RGB image of an app is generated, we adopt ConvNeXt model~\cite{liu2022convnet}, an effective and practical implementation for image-related tasks, to learn the vector profiling the app.
To fit the model, we adjust the size of the images with bilinear interpolation.
Moreover, if some APKs do not contain native code (\ie \textit{.so} files), we set each pixel to zero in the images of the corresponding channel.
Such image-based methods have been shown to be effective in code detection~\cite{sun2021android,wu2022vulcnn,chen2020software}.

\begin{figure}[tb]
\centering\includegraphics[width=\linewidth]{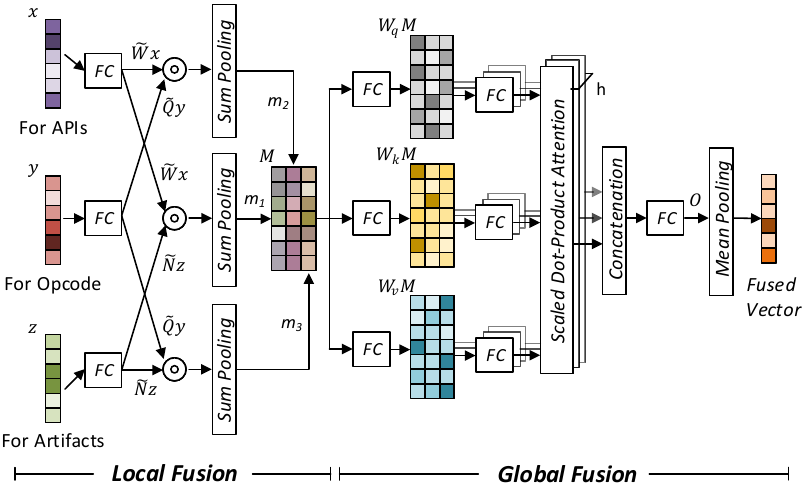}
\caption{A neural network model for fusing feature vectors from three different views with a stepwise strategy. }
\label{vecFuse}
\end{figure}

\subsection{Multi-view Vectorial Fusion}
To combine the strengths of the three views above for describing app behaviors comprehensively, we design a neural network model for multi-view fusion.
It adopts a stepwise strategy performing the vectorial fusion from local to global, as shown in~\Cref{vecFuse}.
With the local fusion, we model the pairwise feature interactions between different views.
With the global fusion, we enhance the representations of the locally-fused vectors by weight based on their contributions to the downstream classification task, and then aggregate the enhanced vectors to depict app behaviors precisely.

In the first step, we aggregate any two of the three vectors from different views with the multi-modal factorized bilinear (MFB) pooling approach~\cite{yu2017multi} respectively.
As a typical approach, MFB makes good performances for fusing multi-view features in both efficiency and effectiveness.
For example, given two feature vectors from different views, \eg, $x \in \mathbb{R}^{d}$ for the view of APIs and $y \in \mathbb{R}^{e}$ for the view of opcode, we fuse them into a single value by
\begin{align}\label{eq1}
    m_{ij} &= \mathds{1}^{T} (W_{j}^{T}x \circ Q_{j}^{T}y)
\end{align}
, where $k$ is the latent dimensionality of the projection matrices $W_{j} \in \mathbb{R}^{d \times k}$ and $Q_{j} \in \mathbb{R}^{e \times k}$,  $\circ$ is the element-wise multiplication of the two vectors, $\mathds{1} \in \mathbb{R}^k$ is an all-one vector.

To obtain the output feature $m_{i} \in \mathbb{R}^{o}$ based on \Cref{eq1}, the weights to be learned are two three-order tensors $W = [W_{1}, W_{2},...,W_{o}] \in \mathbb{R}^{d \times k \times o}$ and $Q = [Q_{1}, Q_{2},...,Q_{o}] \in \mathbb{R}^{e \times k \times o}$ accordingly.
To facilitate the subsequent calculations, we reformulate $W$ and $Q$ as 2-D matrices $\tilde{W} \in \mathbb{R}^{d \times ko}$ and $\tilde{Q} \in \mathbb{R}^{e \times ko}$ respectively with reshape operations.
According to \Cref{eq1}, $M$ is calculated by
\begin{align}
    m_{i} &= \text{SumPooling}(\tilde{W}^{T}x \circ \tilde{Q}^{T}y, k)
\end{align}
, where the function \text{SumPooling}$(t, k)$ means using a one-dimensional non-overlapped window with the size $k$ to perform sum pooling over $t$.
Similarly, the local fusion processes for $x$ and $z$, and $y$ and $z$ are made as depicted above. 

As shown in the left of \Cref{vecFuse}, the local fusion can be implemented by combining some commonly-used layers, \eg, fully-connected, element-wise multiplication.
As explained in the literature~\cite{yu2017multi}, a dropout layer is added after the element-wise multiplication layer to prevent over-fitting.
To optimize convergence of the model, the power normalization and L2 normalization layers are appended after MFB output.

\begin{table}[t]
\caption{The detailed parameters used in our DNN-based Android Malware Classifier.}
\label{tab:para}
\resizebox{\linewidth}{!}{
\begin{tabular}{|c|c|c|c|}
\hline
\textbf{Designed Layer} & \textbf{Dimension} & \multicolumn{1}{l|}{\textbf{Activation Function}} & \textbf{Using Dropout} \\ \hline
Input Layer & 1024 & None & No \\ \hline
Hidden Layer1 & 512 & ReLU & Yes \\ \hline
Hidden Layer2 & 256 & ReLU & Yes \\ \hline
Hidden Layer3 & 128 & ReLU & Yes \\ \hline
Hidden Layer4 & 64 & ReLU & Yes \\ \hline
Output Layer & 2 & Softmax & No \\ \hline
\end{tabular}
}
\end{table}

In the second step, we apply a multi-head self-attention mechanism~\cite{vaswani2017attention} to adjust the representations of all the locally-fused results based on their importance for addressing the specified task.
As shown in the right of \Cref{vecFuse}, $Q = W_{q}M$, $K = W_{k}M$ and $V = W_{v}M$ are firstly produced, where $W_{q} \in \mathbb{R}^{u_{q} \times o}$, $W_{k} \in \mathbb{R}^{u_{k} \times o}$ and $W_{v} \in \mathbb{R}^{u_{v} \times o}$ are the projection matrices, $M$ = [$m_{1}$, $m_{2}$, $m_{3}$] is a vertical concatenation of the locally-fused vectors.
Then the output result for an attention head $head_{i} \in \mathbb{R}^{p_{v}}$ is calculated by
\begin{equation}
    head_{i} = \text{Self-Attention}(W_{i}^{Q}Q, W_{i}^{K}K, W_{i}^{V}V)
\end{equation}
, where $W_{i}^{Q} \in \mathbb{R}^{p_{q} \times u_{q}}$, $W_{i}^{K} \in \mathbb{R}^{p_{k} \times u_{k}}$ and $W_{i}^{V} \in \mathbb{R}^{p_{v} \times u_{v}}$ are the projection matrices.
After that, the output of the multi-head self-attention $O \in \mathbb{R}^{p_{o}}$ is calculated by
\begin{equation}
    O = W^{O}\text{Concat}(head_{1}, head_{2}, ..., head_{h})
\end{equation}
, where $W^{O} \in \mathbb{R}^{p_{o} \times hp_{v}}$ is the projection matrix, and $h$ is the number of attention heads.


Finally, we use the mean pooling to produce the fused vector by compressing the representation results from three views.

\subsection{Malware Classification}
To this end, the fused vectors for apps are used to train a typical DNN model for the malware classification task.
The parameters used in the model are listed in \Cref{tab:para}.
Detailedly, we use ReLU as the activation function in each fully-connected layer.
To avoid the overfitting problem, the dropout strategy is applied in our model, where  the dropout rate is set as 0.2, which is generally used in the typical DNN based models~\cite{kim2018multimodal}.
The output of the last fully-connected layer is process by the Softmax function to obtain the predicted result.
During the learning process, our DNN model is tuned to minimize the value of the loss function, \ie, the cross-entropy function as
\begin{equation}
    J(L, L') = -\sum_{i=1}^{2} L^{i} log L'^{i} + (1 - L^{i})log(1-L'^{i})
\end{equation}
, where $L$ denotes the label of input vector, and $L'$ denotes the predicted label result, $i$ indicates the real categories of the input apps, \ie, malicious or benign.

%% file: evaluations.tex
\section{Experimental Evaluation} \label{evaluation}

To evaluate the effectiveness of LensDroid, we seek to answer the following questions:
\begin{itemize}
    \item \textbf{RQ1:} Does \oursystem detect Android malware better than representative baselines? Does \oursystem possess the capacity for transferability?
    \item \textbf{RQ2:} Whether the fusion of multi-view information in \oursystem enhances the detection of Android malware? 
    Can the three views complement each other? 
    \item \textbf{RQ3:} What are appropriate configurations for the software visualization technique of each view?
    \item \textbf{RQ4:} How effective is the multi-view fusion model in LensDroid? What is the runtime overhead of the model for detecting Android malware?
\end{itemize}

\subsection{Experimental Setup}\label{setup}
\subsubsection{Implementation} We implement a prototype of \oursystem. Specifically, we generate abstract API callgraphs of apps based on the off-the-shelf interfaces of FlowDroid~\cite{DBLP:conf/pldi/ArztRFBBKTOM14}.
We leverage APKTool~\cite{APKTool} to produce \textit{smali} code from \textit{.dex} files in apps. We use Androguard~\cite{Androguard} to extract three kinds of binary files in apps and convert the files to the corresponding RGB images.
The graph generation and learning processes are based on Deep Graph Library and Pytorch respectively.
The experiments are conducted on a server with Intel(R) Xeon(R) Gold 6240 CPU, NVIDIA GeForce RTX 3090 GPU, 128GB memory and Ubuntu
18.04.6 LTS (64 bit).

\begin{table}[t]
\caption{The Number of Apps in Different Categories of Our Datasets.}
\label{tab:datasetNum}
\centering
\resizebox{\linewidth}{!}{
\begin{tabular}{|c|c|c|c|c|c|c|}
\hline
\textbf{Category} & \textbf{2010-2012} & \textbf{2018} & \textbf{2019} & \textbf{2020} & \textbf{2021} & \textbf{2022} \\ \hline
Malware               & 8622               & 3800          & 3900          & 3910          & 3600          & 1600                \\ \hline
Benignware            &      4893    & 4700          & 4700          & 4900          & 4700          & 1840           \\ \hline
\end{tabular}
}
\end{table}


\subsubsection{Datasets}\label{subsubsec:dataset}
To ensure authenticity and reliability of our statistics and the experimental results, as shown in \Cref{sec:motivation}, we collect 51165 samples from AndroZoo and Drebin.
As listed in \Cref{tab:datasetNum}, the collected samples span multiple years.
Since it is inefficient to collect of samples from AndroZoo in the last two years, we randomly gather the samples whose appearance timestamps are within the range of 2010-2012 and 2018-2022.
We regard a sample as malicious if it is flagged by more than 2 antivirus engines, and regard a sample as benign if fewer than 1 engine reports it~\cite{2020Measuring}.
Note that all gathered samples are analyzed by FlowDroid, ApkTool and Androguard without errors and interruptions.

\subsubsection{Baselines} To evaluate our effectiveness, we select five representative Android malware detection tools with different key techniques, each of which is related to \oursystem:
\begin{itemize}
    \item Drebin~\cite{arp2014drebin} is a typical framework that detects an app by collecting a wide range of features, \eg, used hardware, API calls and permissions from the manifest and dex code, and then trains a SVM-based classifier.
    \item Drebin-DL~\cite{grosse2017adversarial} uses the same features sets as Drebin but adopts DNN as the key algorithm to do classification.
    \item MaMaDroid~\cite{onwuzurike2019mamadroid} abstracts each API on call graph with its family or package name to build a first-order Markov chain, and then uses pairwise transition probabilities as the feature vector. We use Random Forests for the downstream malware classification in the experiments.
    \item LBDB~\cite{sun2021android} converts \textit{.dex}, \textit{.so} and \textit{.xml} files of an app into three grayscale images, extracts features from the images by CNN, and generates the representation vector for the app by concatenating the three feature vectors. It then uses DNN to identify Android malware.
    \item N-opcode~\cite{kang2016n} analyzes opcode in apps with N-gram sequences, considering their frequencies and binary counts.
    As the source code is unavailable, we re-implement it from published literature. To avoid dimension inflation of feature vectors, we experiment with N taking values of 2, 3 and 4, and set N as 3 for its superior results.
\end{itemize}

\subsubsection{Evaluation Metrics}
We define true positives as correctly classified malware, false positives as misclassified benignware, true negatives as correctly classified benignware, and false negatives as misclassified malware.
We evaluate with four metrics, including Precision, Recall, Accuracy and F$_{1}$-score.



\begin{table}[t]
\centering
\caption{Comparison of LensDroid and Baselines on Whole Datasets.}\label{tab:whole}
\resizebox{0.95\linewidth}{!}{
\begin{tabular}{|c|c|c|c|c|}
\hline
\textbf{Technique} & \textbf{Precision} & \textbf{Recall} & \textbf{Accuracy} & \textbf{F$_{1}$-score}    \\ \hline
Drebin         & 96.5\%             & 97.9\%          & 97.0\%            & 0.972          \\ \hline
Drebin-DL      & 97.0\%             & 97.9\%          & 97.3\%            & 0.974          \\ \hline
MaMaDroid      & 95.0\%             & 92.8\%          & 95.7\%            & 0.939          \\ \hline
LBDB           & 89.8\%             & 94.0\%          & 89.5\%            & 0.918          \\ \hline
N-opcode       & 95.8\%             & 85.3\%          & 90.6\%            & 0.903          \\ \hline 
\oursystem      & \textbf{98.1\%}    & \textbf{98.5\%} & \textbf{98.3\%}   & \textbf{0.983} \\ \hline
\end{tabular}}
\end{table}

\subsection{RQ1: Effectiveness on Android Malware Detection}
\subsubsection{Overall Effectiveness}\label{wholeData} To validate the detection effectiveness of \oursystem and the baselines, we randomly select 80\% of apps from each of our datasets to form the training set, and test on the rest of the apps. 
The process above is run for 5 times, each time using a different subset for testing, and we calculate the average for each metric as the results.
Due to the long time used to train for an epoch (19 hours on average), we train our model for a maximum of 50 epochs, and use an early stopping strategy with the patience step of 20.

The experimental results are shown in \Cref{tab:whole}.
From the table, we can see that \oursystem outperforms all other baselines in Precision, Recall, Accuracy and F$_{1}$-score.
Specifically, the results of Drebin and Drebin-DL are better than the results of MaMaDroid, LBDB and N-opcode.
After further analysis, we find that the detections of Drebin and Drebin-DL rely on a variety of features from \textit{AndroidManifest.xml} and disassembled code, which help ensure their effectiveness.
MaMaDroid, LBDB and N-opcode focus on only a part of aspects of Android apps, \ie, the sequence of API calls, the contents of three types of binary files and the occurrence or frequency of opcode subsequences respectively, so their detection results are worse than Drebin and Drebin-DL.
In comparison, \oursystem comprehensively assess apps from three views including behavioral sensitivities, operational contexts and supported environments.
Moreover, unlike Drebin and Drebin-DL whose features are organized in sets of strings, \oursystem uses an end-to-end model to automatically learn high-level features that are not inherently linked from the three views, which can express more sufficient semantics.

\begin{table}[t]
\caption{The Comparison of LensDroid and Five Selected Baselines in Three Representative Scenarios.}\label{tab:multiscenrio}
\centering
\resizebox{\linewidth}{!}{
\begin{tabular}{|c|cccc|c|}
\hline
\multirow{2}{*}{\textbf{Technique}} & \multicolumn{4}{c|}{\textbf{App Evolution}} & \multicolumn{1}{c|}{\textbf{Zero Day}} \\ \cline{2-6} 
 & \multicolumn{1}{c|}{\textbf{AUT(p)}} & \multicolumn{1}{c|}{\textbf{AUT(r)}} & \multicolumn{1}{c|}{\textbf{AUT(a)}} & \textbf{AUT(f$_{1}$)} & \multicolumn{1}{c|}{\textbf{Accuracy}} \\ \hline
Drebin & \multicolumn{1}{c|}{92.4\%} & \multicolumn{1}{c|}{85.3\%} & \multicolumn{1}{c|}{90.3\%} & 0.876 & \multicolumn{1}{c|}{24.1\%} \\ \hline
Drebin-DL & \multicolumn{1}{c|}{92.4\%} & \multicolumn{1}{c|}{84.9\%} & \multicolumn{1}{c|}{90.1\%} & 0.873 & \multicolumn{1}{c|}{86.0\%}  \\ \hline
MaMaDroid & \multicolumn{1}{c|}{51.4\%} & \multicolumn{1}{c|}{87.5\%} & \multicolumn{1}{c|}{85.4\%} & 0.548 & \multicolumn{1}{c|}{70.8\%}  \\ \hline
LBDB & \multicolumn{1}{c|}{88.0\%} & \multicolumn{1}{c|}{71.1\%} & \multicolumn{1}{c|}{90.2\%} & 0.786 & \multicolumn{1}{c|}{33.4\%} \\ \hline
N-opcode & \multicolumn{1}{c|}{\textbf{96.7\%}} & \multicolumn{1}{c|}{82.6\%} & \multicolumn{1}{c|}{91.1\%} & 0.879 & \multicolumn{1}{c|}{30.0\%} \\ \hline
LensDroid & \multicolumn{1}{c|}{84.4\%} & \multicolumn{1}{c|}{\textbf{99.2\%}} & \multicolumn{1}{c|}{\textbf{92.9\%}} & \textbf{0.897} & \multicolumn{1}{c|}{\textbf{93.2\%}}  \\ \hline
\end{tabular}
}
\end{table}

\subsubsection{Method Transferability}
To verify transferability (\ie, detectability to unknown types of apps) of our work, we perform two experiments on evolved apps and zero-day malware.
Experimental procedure is same as depicted in \Cref{wholeData}.

\textit{\textbf{Adaptation for App Evolution.}} DL-based Android malware detectors face a problem of aging as the app evolution due to the update of Android versions~\cite{zhang2020enhancing}.
In the experiment, we treat the samples in 2018 as the training set and the samples whose appearance timestamps are within the range of 2019-2022 as four testing sets respectively.
Considering that the appearance timestamps of the apps in the training set are earlier than the appearance timestamps of the apps in the testing set, we experiment to figure out whether a detector can identify evolved Android malware.
We use Area Under Time (AUT) to measure a classifier's robustness to time decay~\cite{pendlebury2019tesseract}:
\begin{equation}
AUT(f,N)= \frac{1}{N-1} \sum_{k=1}^{N-1} \frac{f(k+1) + f(k)}{2}
\end{equation}
, where $f$ is our evaluation metric, $N$ is the number of test slots, and $f(k)$ is the evaluation metric generated at time $k$.
N is set as 12 months in our experiment and thereby omitted from AUT expressions
 in subsequent sections.
An AUT metric that is closer to 1 means better performance over time.

The experimental results are list in \Cref{tab:multiscenrio}, where LensDroid outperforms all baselines in AUT(r), AUT(a) and AUT(f$_{1}$), but has lower AUT(p) than the baselines.
Specifically, Drebin, Drebin-DL, LBDB and N-opcode are better than \oursystem in Precision-related metric AUT(p).
Considering that the Precision of \oursystem is higher than all baselines on the whole datasets in \Cref{tab:whole}, we can know that the Precision of \oursystem is relatively poor on the datasets from few years.
Nonetheless, AUT(r) of \oursystem, \ie, 99.2\%, is much higher than the results of all baselines, which means that \oursystem focuses on not missing malware samples as much as possible over time.
Owing to the complementary nature of the used views in capturing invariant factors of apps, the performance of \oursystem is better than each of the selected baselines in terms of the comprehensive metrics, \ie, AUT(a) and AUT(f$_{1}$).
Moreover, it is interesting that N-opcode has the best overall performance of all baselines, which shows that opcode-related features are rather stable over time than API-related, content-related and hybrid features in our experiment.

\begin{figure}[tb]
\centering\includegraphics[width=\linewidth]{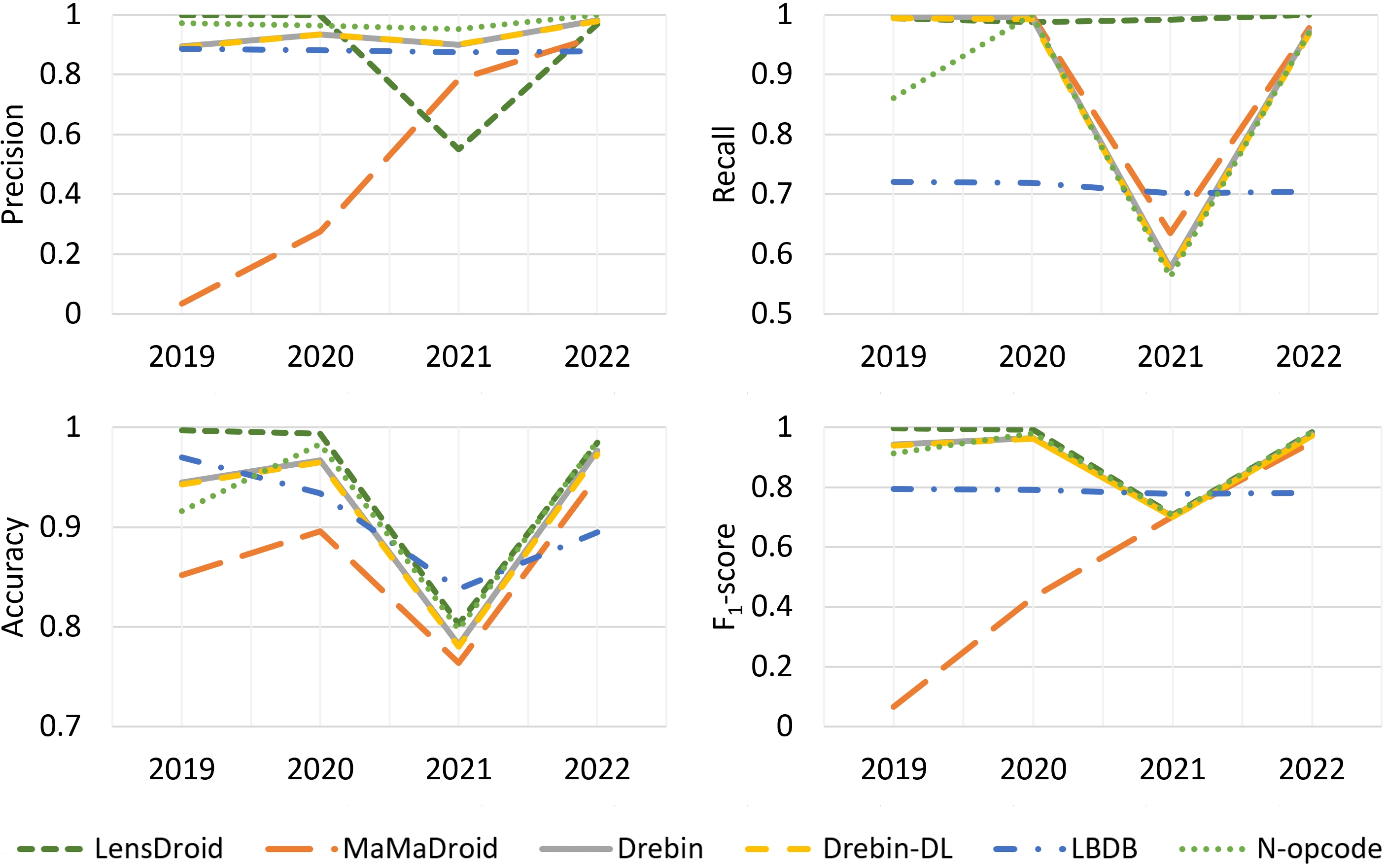}
\caption{The performance metrics over 4 test slots of total 48 months.}
\label{fig:appevo}
\end{figure}

We then plot the change trend of the evaluation metrics for each tool over time in \Cref{fig:appevo}.
Overall, the metrics of most tools fluctuate significantly and then drop sharply in 2021.
Specifically, we can see that the reason for the lower AUT(p) of \oursystem is the poor detection performance on the apps in 2021.
Nevertheless, the effectiveness of \oursystem for detecting the apps in 2022 has picked up again.
Therefore, we suggest that a detection model should be retrained with newer app data after two years of use in practice, which is similar to the advice in MsDroid~\cite{he2022msdroid}.
In comparison, the Precision of N-opcode is stable over time.
The gradual improvement of the F$_1$-score for MaMaDroid indicates that the sequence of API calls can be regarded as potential clues to
label malware samples over time.
The analysis results above are the guidance for users to select suitable tools on demand.

\textit{\textbf{Identification of Zero-day Malware.}}
Zero-day malware refers to a kind of unknown malicious apps whose signatures are not available
before.
To construct the required experimental environment, we first use the malware samples from the Drebin dataset and the benign samples whose timestamps fall within the Drebin date range as our training set.
To distinguish this experiment from app evolution, we then collect malware samples from AndroZoo, which are not classified into the families present in the Drebin dataset and have the same timestamps as above, to serve as our testing set.
Here we use AVClass~\cite{sebastian2016avclass} to obtain the families of the samples.
This takes the size of our zero-day malware samples to 3332.

Since all the tested samples are malware in the scenario, we only calculate Accuracy as shown in \Cref{tab:multiscenrio}.
Considering that there is no benignware in the testing set, the accuracy and recall are calculated in the same way.
Therefore, we directly compare the Accuracy with the Recall on the whole dataset in \Cref{tab:whole}.
Specifically, the effectiveness of each tool has declined, especially for Drebin (73.8\%), Drebin-DL (11.9\%), MaMaDroid (22\%), LBDB (60.6\%) and N-opcode (50.3\%).
We conclude that using inappropriate methods to aggregate various information or only relying on the information from one single view is difficult to prevent effectiveness reduction in zero-day malware detection.
In comparison, \oursystem achieves higher accuracy than other tools, where our detection results are only down 5.3\%.
It means that the complementarity of the views helps gain in-depth knowledge and assess maliciousness of unknown apps comprehensively and precisely.


\begin{table}[t]
\caption{Effectiveness of LensDroid on Different Combinations of Views for Apps (V$_1$ = behavioral sensitivities, V$_2$ = operational contexts, V$_3$ = supported environments).}\label{tab:comb}
\centering
\resizebox{0.95\linewidth}{!}{
\begin{tabular}{|c|c|c|c|c|}
\hline
\textbf{View} & \textbf{Precision} & \multicolumn{1}{l|}{\textbf{Recall}} & \textbf{Accuracy} & \textbf{F$_{1}$-score} \\ \hline
V$_1$ & 94.3\% & 95.5\% & 94.8\% & 0.949 \\ \hline
V$_2$ & 94.7\% & 96.8\% & 95.6\% & 0.957 \\ \hline
V$_3$ & 94.9\% & 93.8\% & 94.2\% & 0.944 \\ \hline
V$_1$+V$_2$ & 95.3\% & 97.7\% & 96.4\% & 0.965 \\ \hline
V$_1$+V$_3$ & 95.0\% & 97.1\% & 96.0\% & 0.960 \\ \hline
V$_2$+V$_3$ & 96.5\% & 97.0\% & 96.7\% & 0.967 \\ \hline
V$_1$+V$_2$+V$_3$ & \textbf{98.1\%} & \textbf{98.5\%} & \textbf{98.3\%} & \textbf{0.983} \\ \hline
\end{tabular}}
\end{table}

\subsection{RQ2: Complementarity of Multiple Views} 
\subsubsection{Combinations of Different Views}\label{ablation} To verify the effectiveness of multi-view fusion, we perform ablation experiments on seven combinations of the three views.
Experimental procedure and datasets are the same as depicted in \Cref{wholeData}.

The experimental results are shown in \Cref{tab:comb}, where the views of behavioral sensitivities, operational contexts and supported environments are represented as V$_1$, V$_2$ and V$_3$ accordingly.
From the table, we observe that 
the integration of the three views in \oursystem incorporates semantic information within apps at different levels, thus yielding better detection results.
Specifically, under single view, the Accuracy of LensDroid exceeds 94.2\%, which is higher than the Accuracy of LBDB and N-opcode in \Cref{tab:whole}.
Meanwhile, the F$_1$-score of LensDroid exceeds 0.944, which is higher than the same metric of MaMaDroid, LBDB and N-opcode in \Cref{tab:whole}.
It shows that each of the views is capable of detecting a large number of Android malware in practice.
Moreover, the results of operational contexts are the best of the three, which demonstrates the effectiveness of opcode sequences in distinguishing benignware and malware as before.
When combining any two views, the Accuracy and F$_1$-score of LensDroid exceed 96.0\% and 0.960 respectively, both of which are higher than the same metrics of MaMaDroid, LBDB and N-opcode in \Cref{tab:whole}.
It means that incorporating information from the two views helps identify more Android malware.
Finally, the integration of the information from the three views makes LensDroid determine the maliciousness of apps comprehensively (\ie, 98.3\% Accuracy and 0.983 F$_1$-score).
This verifies the effectiveness of the multi-view fusion of our work.
Therefore, we make the following summaries:

\begin{itemize}
    \item The software visualization technique tailored for each view in our work can help extract insightful semantics to characterize Android malware.
    In essence, the visualized features derived from each view make a positive contribution for revealing hidden maliciousness in apps.
    
    \item The high-level features obtained from different views can complement each other to achieve an effective malware detection.
    As the number of views combined increases, all of LensDroid's detection metrics improve to varying degrees.
    Furthermore, this also illustrates the effectiveness of our multi-view fusion method in aggregating and releasing the potential detectability of the views.
\end{itemize}

\subsubsection{Case Study}
To validate the complementarity of the three views, we choose two apps and show the reasons for classifying them with GNNExplainer~\cite{ying2019gnnexplainer} and Grad-CAM~\cite{selvaraju2017grad}.
The app named \textit{ke.co.ulezi.Ulezi}\footnote{\texttt{MD5: d808a309179783242e2a5e1e5763320d}} is a benignware sample, and the app named \textit{com.androidbox.cwwgnet8}\footnote{\texttt{MD5: 997a62aec35fe1355ebf470baa3a89ae}} is a malware sample.
The two apps are detected by LensDroid correctly but misclassified by all the baselines. 
To demonstrate the effectiveness of LensDroid, we use GNNExplainer to output the edges contributing significantly to the app classification decision on API callgraphs.
We use Grad-CAM to highlight the important regions for the malware identification on opcode-gram-based matrices and binary-transformed images.

\begin{figure}[tb]
\centering\includegraphics[width=\linewidth]{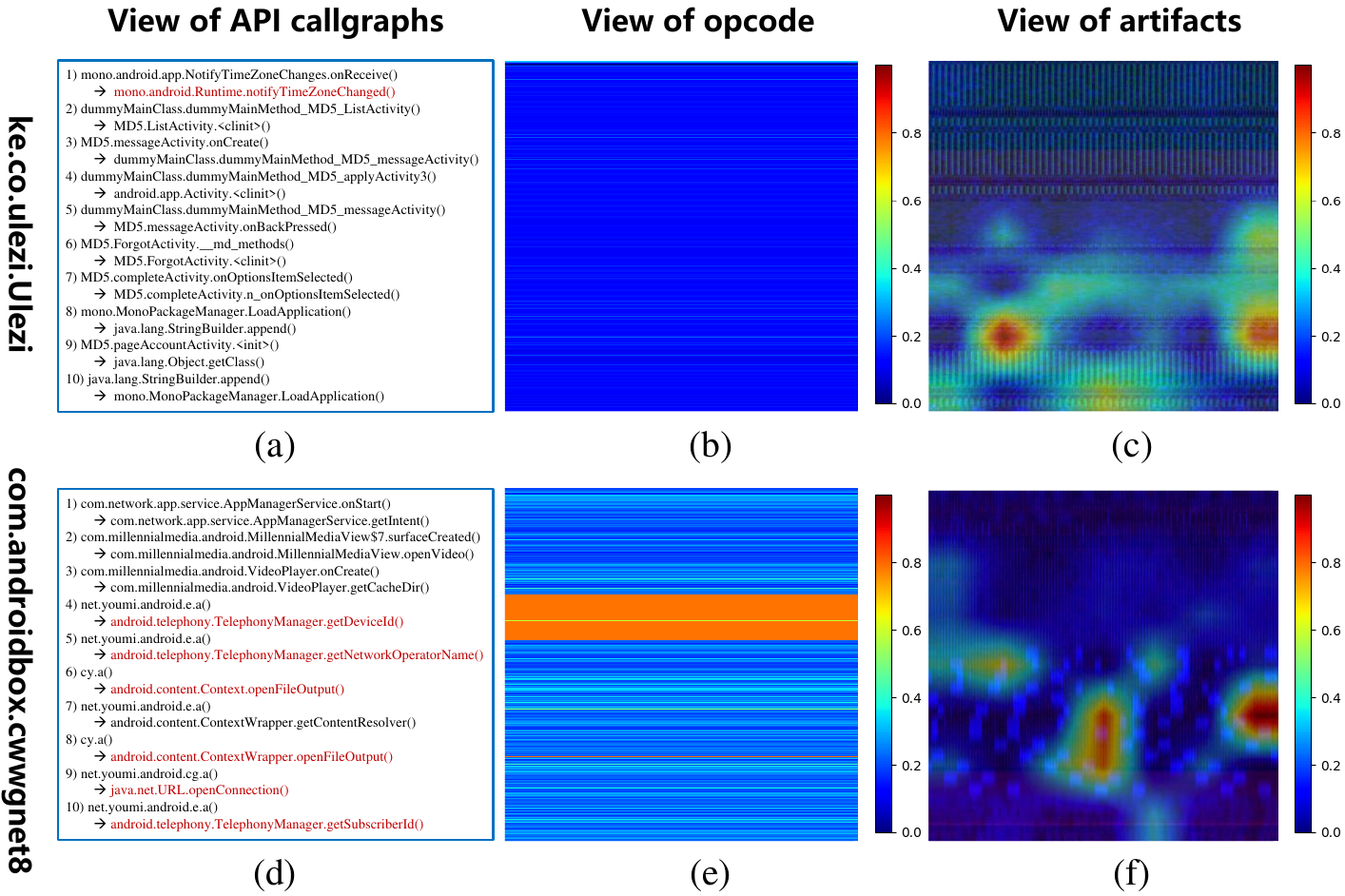}
\caption{The top-10 API call edges for the view of API callgraphs (sensitive APIs are in red), and the heatmaps of the views of opcode and artifacts for two representative apps named \textit{ke.co.ulezi.Ulezi} and \textit{com.androidbox.cwwgnet8}.}
\label{fig:case}
\end{figure}

\Cref{fig:case} shows the top-10 API call edges and heatmaps produced by the aforementioned tools for the two apps.
Based on the figure, we summarize as follows:
\begin{itemize}
    \item The top-10 API call edges between the apps are different. Specifically, the APIs listed in \Cref{fig:case}(a) are insensitive, where most of them are used to manipulate app UIs and only \textit{notifyTimeZoneChanged()} can be used to trigger malicious behaviors.
    In comparison, many sensitive APIs are shown in \Cref{fig:case}(d). 
    For example, the library named \textit{youmi} calls \textit{getDeviceId()}, \textit{getNetworkOperatorName()}, \textit{getSubscriberId()} to collect user privacy and uses \textit{openConnection()} to create a remote connection. Another library named \textit{cy} invokes \textit{openFileOutput()} to open a specified file.
    The above analysis supports LensDroid in making the correct classification.

    \item The difference of the opcode-gram-based matrices between the apps is significant. Specifically, there are almost no highlighted regions on the heatmap of the matrix of the benignware sample in \Cref{fig:case}(b). In comparison, there are obvious regions highlighted on the heatmap of the malware sample in \Cref{fig:case}(e). This means that LensDroid can discover key clues based on the view of opcode to distinguish between malware and benignware.
    \item There are some similarities between the highlighted regions on binary-transformed images of the apps. 
    Specifically, the relative positional relationship of red areas in \Cref{fig:case}(c) is somewhat similar to that in \Cref{fig:case}(f).
    However, the size and shape of the red areas in the two subfigures are different.
    This means that LensDroid can find suspicious parts based on the view of artifacts, while misclassification may be made using only this view.
\end{itemize}

Evaluating maliciousness of an app from multiple views can obtain diverse insights, which delineate the characteristics of the app at different levels.
Owing to the semantic complementary among the three views, LensDroid is capable of automatically extracting crucial information from the insights, thereby generating a comprehensive evaluation criterion.
Therefore, even though it is not enough to distinguish the two apps from one view (\eg, \Cref{fig:case}(c) and \Cref{fig:case}(f)), LensDroid can still make correct decisions with insightful information.


\begin{table}[t]
\caption{Experimental Results for Adjusting Configurations of Software Visualization Techniques under Different Views.}\label{tab:scene}
\centering
\resizebox{0.95\linewidth}{!}{
\begin{tabular}{|ccccc|}
\hline
\multicolumn{1}{|c|}{\textbf{Configuration}} & \multicolumn{1}{c|}{\textbf{Precision}} & \multicolumn{1}{c|}{\textbf{Recall}} & \multicolumn{1}{c|}{\textbf{Accuracy}} & \textbf{F$_1$-score} \\ \hline
\rowcolor{gray!20}
\multicolumn{5}{|c|}{\textbf{API Abstraction}} \\ \hline
\multicolumn{1}{|c|}{Randomly} & \multicolumn{1}{c|}{89.0\%} & \multicolumn{1}{c|}{91.4\%} & \multicolumn{1}{c|}{90.0\%} & 0.902 \\ \hline
\multicolumn{1}{|c|}{Ours} & \multicolumn{1}{c|}{\textbf{94.3\%}} & \multicolumn{1}{c|}{\textbf{95.5\%}} & \multicolumn{1}{c|}{\textbf{94.8\%}} & \textbf{0.949} \\ \hline
\rowcolor{gray!20}
\multicolumn{5}{|c|}{\textbf{Length of the Sliding Window}} \\ \hline
\multicolumn{1}{|c|}{\textit{Length} = 1} & \multicolumn{1}{c|}{79.2\%} & \multicolumn{1}{c|}{97.3\%} & \multicolumn{1}{c|}{88.2\%} & 0.873 \\ \hline
\multicolumn{1}{|c|}{\textit{Length} = 2} & \multicolumn{1}{c|}{88.8\%} & \multicolumn{1}{c|}{96.0\%} & \multicolumn{1}{c|}{92.4\%} & 0.923 \\ \hline
\multicolumn{1}{|c|}{\textit{Length} = 3} & \multicolumn{1}{c|}{89.9\%} & \multicolumn{1}{c|}{96.8\%} & \multicolumn{1}{c|}{93.3\%} & 0.932 \\ \hline
\multicolumn{1}{|c|}{\textit{Length} = 4} & \multicolumn{1}{c|}{\textbf{94.7\%}} & \multicolumn{1}{c|}{96.8\%} & \multicolumn{1}{c|}{\textbf{95.6\%}} & \textbf{0.957} \\ \hline
\multicolumn{1}{|c|}{\textit{Length} = 5} & \multicolumn{1}{c|}{90.0\%} & \multicolumn{1}{c|}{97.1\%} & \multicolumn{1}{c|}{93.5\%} & 0.934 \\ \hline
\multicolumn{1}{|c|}{\textit{Length} = 6} & \multicolumn{1}{c|}{90.7\%} & \multicolumn{1}{c|}{\textbf{97.6\%}} & \multicolumn{1}{c|}{94.1\%} & 0.940 \\ \hline
\rowcolor{gray!20}
\multicolumn{5}{|c|}{
\textbf{Artifact Processing}} \\ \hline
\multicolumn{1}{|c|}{Undenoising} & \multicolumn{1}{c|}{90.5\%} & \multicolumn{1}{c|}{\textbf{95.1\%}} & \multicolumn{1}{c|}{92.7\%} & 0.927 \\ \hline
\multicolumn{1}{|c|}{Denoising} & \multicolumn{1}{c|}{\textbf{94.9\%}} & \multicolumn{1}{c|}{93.8\%} & \multicolumn{1}{c|}{\textbf{94.2\%}} & \textbf{0.944} \\ \hline
\end{tabular}}
\end{table}

\subsection{RQ3: Configurations for Each View} \label{sec:config}
To take full advantage of the software visualization techniques, we perform three experiments, each of which adjusts configurations of the technique in a single view and decides appropriate configurations.
Experimental procedure and datasets are the same as described in \Cref{ablation}.

\subsubsection{View of Behavioral Sensitivities} \oursystem makes the abstractions for APIs to describe their sensitivities.
To demonstrate the effectiveness of this strategy, referring to the treatment of the representative work~\cite{kipf2016semi}, we randomly initialize node representation vectors for APIs in callgraphs, and then compare the experimental results with our work.
 

The comparison results are depicted in Lines 3-4 of \Cref{tab:scene}, where the API abstraction method works better than random initialization.
Specifically, randomly initializing API nodes in callgraphs ignores the variations in the sensitivity strength among various APIs, consequently leading GCN to integrate imprecise API sensitivity information.
That results in biased vector representations.
In comparison, the sensitivity representation of APIs distinguishes the sensitivities of different APIs in callgraphs and helps GCN perform the propagation and aggregation of featural information of APIs accurately.

\subsubsection{View of Operational Contexts} To explore a suitable length of the sliding window in \Cref{sec:context}, we experiment by adjusting \textit{Length} from 1 to 6 for extracting different opcode subsequences and detecting Android malware in our datasets.

The experimental results are listed in Lines 5-11 of \Cref{tab:scene}, in which Precision, Accuracy and F$_1$-score are all higher than others when \textit{Length} is set as 4.
This implies that extracting opcode subsequences by a window of length 4 enables effective encapsulation of the contextual information that is crucial for detecting Android malware.
In comparison, setting \textit{Length} to 6 can get a higher recall, but the comprehensive metrics (\ie, Accuracy and F$_{1}$-score) under the setting is lower.
Therefore, we set \textit{Length} as 4 in \oursystem.



\begin{figure}[tb]
\centering\includegraphics[width=\linewidth]{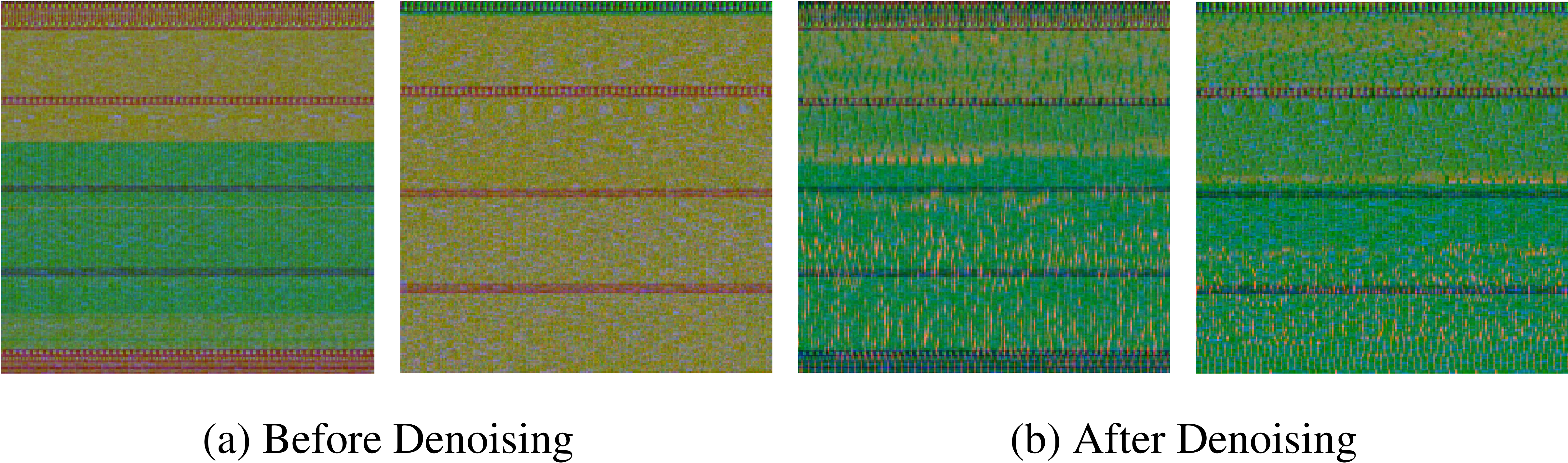}
\caption{The comparison for visualizing two representative apps from the view of supported environments before and after denoising (the left one is \textit{ai.lighten.lighten} and the right one is \textit{bhygzs.top.iotykcar} in each subfigure).}
\label{fig:image}
\end{figure}

\subsubsection{View of Supported Environments} To validate that the noise removal strategy on \textit{.dex}, \textit{.so} and \textit{.xml} files facilitates Android malware detection, we compare the detection results of \oursystem with and without denoising.

The comparison results are listed in Lines 12-14 of \Cref{tab:scene}, where Precision, Accuracy and F$_1$-score are improved by 4.4\%, 1.5\% and 0.017 respectively, but Recall is declined slightly.
In terms of comprehensive metrics,
the indexing and metadata information interferes with the judgment of maliciousness of app behaviors to some extent.
It is practical to perform denoising on the contents of artifacts in apps.


To assess the impact of the denoising, we compare two apps: \textit{ai.lighten.lighten}\footnote{\texttt{MD5: 615a0045dabd4dc009234e14baecc9ad}} and \textit{bhygzs.top.iotykcar}\footnote{\texttt{MD5: bef5785763cfdf06ef83227f0c020040}}.
\Cref{fig:image} illustrates the RGB images with and without denoising.
Both apps are malware, yet their visualized RGB images differ markedly.
Manually analyzing their \textit{.dex} files reveals extensive indexing and metadata unrelated to malicious app behaviors.
Consequently, the left and the right in \Cref{fig:image}(a) show clear distinctions.
The denoising alters the RGB images.
The RGB images of the same app before and after denoising (\eg, the left parts in \Cref{fig:image}(a) and \Cref{fig:image}(b)) are different visibly.
Moreover, the denoised RGB images of both apps (\ie, the left and the right in \Cref{fig:image}(b)) exhibit clear similarities, helping them be classified as the same app type.



\subsection{RQ4: Performance of Our Model}

\begin{table}[t]
\caption{The Comparison of the Model We Used and the Alternative Models in Effectiveness of Android Malware Detection.}\label{tab:concat}
\centering
\resizebox{0.96\linewidth}{!}{
\begin{tabular}{|c|c|c|c|c|}
\hline
\textbf{Model} & \textbf{Precision} & \textbf{Recall} & \textbf{Accuracy} & \textbf{F$_1$-score}    \\ \hline
w/o Self-Attn & \multicolumn{1}{c|}{97.1\%}  & \multicolumn{1}{c|}{98.2\%}  &  \multicolumn{1}{c|}{97.9\%}   &0.976   \\ \hline
w/o MFB  & \multicolumn{1}{c|}{96.4\%}    & \multicolumn{1}{c|}{97.8\%} &\multicolumn{1}{c|}{97.4\%}     &0.971   \\ \hline
Add Only   &  \multicolumn{1}{c|}{88.4\%}   &  \multicolumn{1}{c|}{97.3\%} &\multicolumn{1}{c|}{94.4\%} &0.934    \\ \hline
Ours        &  \multicolumn{1}{c|}{\textbf{97.9\%}}    & \multicolumn{1}{c|}{\textbf{98.9\%}}  &\multicolumn{1}{c|}{\textbf{98.5\%}}    &\textbf{0.984}    \\ \hline
\end{tabular}}
\end{table}

\begin{figure*}[t]
\centering
\includegraphics[width=\textwidth]{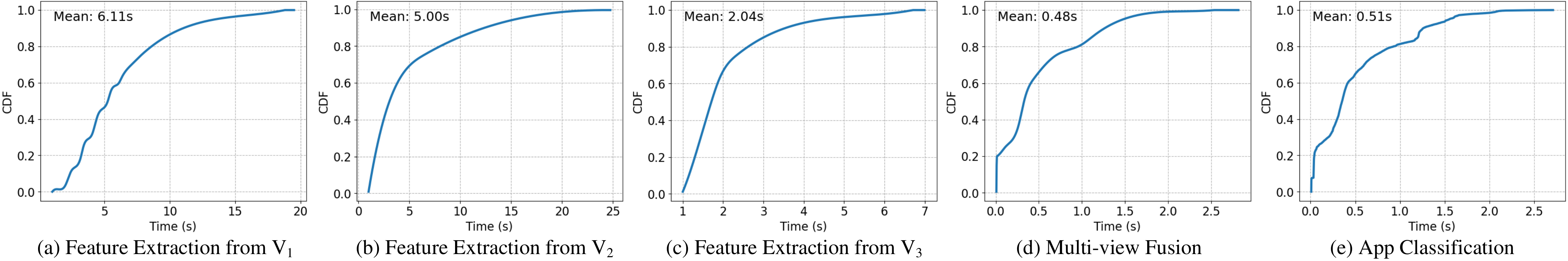}
\caption{The Cumulative Distribution Function (CDF) of the runtime overhead of \oursystem on five phases (V$_1$ = behavioral sensitivities, V$_2$ = operational contexts, V$_3$ = supported environments).}
\label{fig:image5}
\end{figure*}

\subsubsection{Effectiveness on Multi-view Fusion Model} 
To ensure experimental efficiency, we randomly select 60\% of the apps from our datasets, which encompass over 30K apps across various years, to train a model for 10 epochs.
Due to the computational constraints of our server as depicted in \Cref{setup}, training our model for an epoch with the chosen dataset is estimated to take over 9 hours on average.
Multiple models need to be trained for comparison, so we train a model with less epochs.
As observed in the experiments before, the performance of our model tends to stabilize basically after training for 10 epochs.
Other experimental environment and procedure are the same as depicted in \Cref{wholeData}.

To validate if the multi-view fusion model in LensDroid facilitates Android malware detection, we compare the effectiveness of \oursystem when using the original and alternative modules.
We construct three alternative models based on the techniques in \oursystem:

\begin{itemize}
    \item It preserves MFB pooling but replaces multi-head self-attention with vector averaging (\ie, w/o Self-Attn). 
    \item It retains multi-head self-attention but replaces MFB pooling with vector addition (\ie, w/o MFB).
    \item The module only leverages vector addition to implement the multi-view fusion (\ie, Add Only).
\end{itemize}

\Cref{tab:concat} presents the experimental results, showing that our original model surpasses the alternative models in detection effectiveness.
Specifically, vector addition performs the worst with 94.4\% Accuracy and 0.934 F$_1$-score.
Both MFB pooling and multi-head self-attention mechanism improve detection results.
The former achieves 97.9\% Accuracy and 0.976 F$_1$-score, and the latter reaches 97.4\% Accuracy and 0.971 F$_1$-score.
The two modules contribute to the detection, and have different characteristics in multi-view fusion.
Therefore, our model peaks at 98.5\% Accuracy and 0.984 F$_1$-score.

To the best of our knowledge, it is the first to combine MFB pooling with self-attention mechanism to detect Android malware, though the two techniques are not novel.
Experimental results validate its effectiveness.
We will further improve the model in the future, which is orthogonal to this work.

\subsubsection{Runtime Overhead of Our Model}
We evaluate the runtime overhead of \oursystem with entire datasets.
Given an app, LensDroid has five main phases, including feature extractions for three views, multi-view fusion, and app classification.

\Cref{fig:image5} plots the runtime overhead of \oursystem in different phases.
Among the three phases of feature extractions, the process on the view of behavioral sensitivities takes the longest time, averaging 6.11 seconds, and 96.73\% of apps are analyzed within 15 seconds.
In comparison, the process on the view of supported environments is the quickest, averaging 2.04 seconds, and 96.85\% of apps are analyzed within 5 seconds.
The multi-view fusion process is fastest of all the phases, taking only 0.48 seconds on average, and 95.13\% of apps complete fusion in 1.5 seconds.
It indicates that \oursystem can efficiently fuse feature vectors from the three views.
In the final phase, \oursystem leverages a standard DNN model for app classification, which takes about 0.51 seconds on average, and 98.49\% of apps complete classification in 2 seconds.

Overall, the runtime overhead of \oursystem is reasonable.
In practice, \oursystem's performance can be further optimized by increasing the computational power of our server, parallelizing feature extraction tasks and other improvements.


%% file: discussion.tex

\section{Limitation and Discussion}


\subsection{Granularity of Behavioral Analysis} \label{grad}
Different from existing fine-grained analysis schemes~\cite{arp2014drebin,DBLP:conf/pldi/ArztRFBBKTOM14,DBLP:conf/icse/OcteauLDJM15} for app behavioral details, \oursystem uses a data-driven approach to assess app behaviors.
Specifically, \oursystem leverages software visualization techniques to extract and aggregate high-level features of app behaviors for achieving the precise malware identification, rather than explicitly model and analyze behavioral details, \eg, sensitive data flows, ICC links, JNI usages.
Therefore, compared to the aforementioned schemes, \oursystem is less intuitive in explaining why an app is classified as malware.
We have explored the complementarity of the three views by GNNExplainer and Grad-CAM.
To further improve the interpretability, we plan to locate suspicious factors of malicious app behaviors by analyzing kinds of weights in our model.

Since our work utilizes a data-driven method,
it holds promise for extension to identify specific Android attacks,
\eg, code obfuscation~\cite{gao2023obfuscation} or dynamic code loading~\cite{alhanahnah2020dina}.
Some of the features we employ (\eg, images converted from app code) are weakly resistant to challenging environments~\cite{gao2024comprehensive}.
To cover more types of Android malware, we will attempt to design intermediate representations aligned with practical detection requirements~\cite{gao2023obfuscation,zhang2020enhancing}, or generate more robust features with disentangled representation learning~\cite{hou2021disentangled}. 

\subsection{Robustness of Adopted Tools}
Due to limitations from the well-known tools we adopt, including FlowDroid, Androguard and APKTool, \oursystem may omit behavioral details when visualizing apps from different views. 
Specifically, the three tools are all implemented based on static analysis, hence some inherent defects are inevitable.
For example, they are incapable of handling dynamic mechanisms, \eg, dynamic code loading, Java reflective calls, so some method call relations are missed in the generated callgraphs.
Furthermore, similar to the descriptions of previous works~\cite{pauck2018android,onwuzurike2019mamadroid}, Androguard and APKTool fail to decompile some peculiar Android apps correctly.
In the future, we plan to introduce advanced program analysis, \eg, hybrid analysis~\cite{tsutano2019jitana}, for obtaining more comprehensive app visualized features.

\subsection{Configurability of Each View}

\subsubsection{Generation of Sensitivity Representations}
To provide resilience to API changes and obtain semantic commonalities of them, \oursystem generates and merges sensitivity representations for APIs based on the mapping from a typical work named PScout.
The experimental results validate the effectiveness of our work under this configuration.
To increase the coverage of different APIs in the future, \oursystem supports enriching the mapping with the results of Axplorer~\cite{backes2016demystifying}.

\subsubsection{Length of the Sliding Window}
\oursystem utilizes the sliding window to extract semantics within a specified opcode sequence.
We determine the appropriate value of the length of the sliding window through empirical experiments in \Cref{sec:config}.
In practice, analysts can choose the \textit{Length} experimentally based on their collected datasets.

\subsubsection{Content of Artifacts within APKs}
To achieve a precise Android malware detection, we eliminate contents that are not relevant to app behavioral analysis
from \textit{.dex}, \textit{.xml} and \textit{.so} files.
To satisfy broader detection needs for analysts, we plan to partition the content of the artifacts and rule out unconcerned components automatically based on preprocessed techniques, \eg, code slicing~\cite{hoffmann2013slicing}, 
data-flow tracking~\cite{DBLP:conf/pldi/ArztRFBBKTOM14}. 

%% file: relatedwork.tex

\section{Related Work}

\subsection{Single-view Detections}

\subsubsection{Signature-based Approaches}
VAHunt~\cite{shi2020vahunt} identified app-virtualization-based malware based on the given loading strategies of plugins within apps.
PermPair~\cite{arora2019permpair} detected Android malware by the usage of permission pairs declared in manifest files.
DINA~\cite{alhanahnah2020dina} was a hybrid analysis method that detects malicious IAC activities in dynamically loaded code based on the patterns of reflective/dynamic class loading calls.

Each of the aforementioned methods can only identify a specified type of Android malicious behaviors.
\oursystem mines hidden maliciousness of different kinds of app behaviors comprehensively using DL techniques to fuse high-level features extracted from the three related but distinct views.

\subsubsection{ML-based Approaches}
In the application of traditional ML techniques, 
MaMaDroid~\cite{onwuzurike2019mamadroid} performed malware classification based on Markov chains of the sequences of API calls in apps.
HomDroid~\cite{wu2021homdroid} found Android covert malware by analyzing the homophily of call graphs from social-network-based methods.
Difuzer~\cite{samhi2022difuzer} uncovered suspicious hidden sensitive operations/logic bombs by training an one-class SVM.

With representative learning, 
MsDroid~\cite{he2022msdroid} identified malicious snippets in Android apps with building the behavior subgraph set and performing the graph embedding.
AMCDroid~\cite{liu2023enhancing} detected Android malware based on a homogenous graph generated from the call graph and code statements.
Inspired by the development of DL, some works converted code of given languages, \eg, Android~\cite{sun2021android,he2023resnext+},
into a variety of images and then performed detection tasks with CNNs.

The methods above focus on analyzing different kinds of code from one given view.
To achieve a precise and effective malware detection, \oursystem utilizes strengths of DL techniques in aspects of feature extraction and code representation for depicting intentions of app behaviors from multiple views.

\subsection{Multi-view Detections}
Drebin~\cite{arp2014drebin} built feature vectors by extracting eight types of static information from apps and generated a malware classifier by training a SVM model.
Kim et al.~\cite{kim2018multimodal} extracted existence-based and similarity-based features from apps and then fed the vectors to a multimodal deep neural network model for identifying Android malware.
In these methods, the design of feature vectors relies on considerable research experience and manual effort.
Moreover, the methods perform straightforward combinations of multi-view features, \ie, just concatenating the feature vectors from individual views.
\oursystem automatically extracts high-level features from the three different views and leverages an attention-based deep neural network to exploit the complementarity of the features.

MKLDroid~\cite{narayanan2018multi} used a graph kernel to capture structural and contextual information from dependency graphs and identified malicious code patterns from multiple view.
APK2VEC~\cite{narayanan2018apk2vec} performed the dependency analysis based on the inter-procedural control-flow graph (ICFG) of apps, and then leveraged the semi-supervised multimodal representation learning technique to build app profiles automatically.
These methods leverage multiple views derived from the provided ICFGs to detect Android malware, relying on the semantics inherent in graph structures.
\oursystem relies on not only the information from graph structures but opcode, libraries, configurations, \etc
Similarly, Sun et al.~\cite{sun2021android} detected Android malware by converting \textit{.xml}, \textit{.so} and \textit{.dex} files of apps as the corresponding images, which are also a part of the aspects of our method. 
CorDroid~\cite{gao2023obfuscation} developed an obfuscation-resilient Android malware analysis based on enhanced sensitive function call graphs and opcode-based markov transition matrixes.
CorDroid and LensDroid serve different purposes, where the former focuses on combating code obfuscation, while the latter aims to assess maliciousness hidden in apps comprehensively based on heterogeneous features.
As discussed in \Cref{grad}, our work can be tailored to analyze obfuscated apps in the future by refining the input features.

Some techniques modelled app information under different views as heterogeneous graphs, and then performed representation learning to generate feature vectors~\cite{hou2017hindroid,niu2023gcdroid}.
The schemes depict the relations of Android entities between different apps explicitly.
\oursystem aggregates high-level features that are not inherently linked from three views, so as to depict hidden semantics in app behaviors.

Qiu et al.~\cite{qiu2022cyber} also extracted features from source code, API callgraphs, and \textit{smali} opcode to detect Android malware.
However, the tool and LensDroid use different approaches to extract and fuse features.
Specifically, the feature extraction of the tool is relatively straightforward.
For example, it builds the vectors under source code view by manually selecting essential strings.
Meanwhile, it fuses feature vectors with a basic DNN model.
In comparison, \oursystem produces feature vectors with high-order semantics under different views by suitable DL techniques.
We fuses the vectors based on their contributions to malware identification.
Therefore, \oursystem can unleash detection potentials of the adopted views better.

%% file: conclusion.tex

\section{Conclusion}

We propose and implement \oursystem, a novel technique that detects Android malware by visualizing app behaviors from multiple complementary views.
To thoroughly comprehend the details of apps, we decouple the analysis of app behaviors into three related but distinct views of behavioral sensitivities, operational contexts and supported environments.
We extract high-level features based on abstract API callgraphs, opcode-gram-based matrices and binary-transformed images.
To exploit the complementarity of the views, we design a DNN based model for fusing the visualized features from local to global.
In our comprehensive evaluations, \oursystem outperforms five baselines for detecting Android malware under three practical scenarios.
We validate the complementarity of the views and demonstrate that the multi-view fusion in LensDroid enhances Android malware detection.

%% file: bare_jrnl.bbl
\begin{thebibliography}{10}
\providecommand{\url}[1]{#1}
\csname url@samestyle\endcsname
\providecommand{\newblock}{\relax}
\providecommand{\bibinfo}[2]{#2}
\providecommand{\BIBentrySTDinterwordspacing}{\spaceskip=0pt\relax}
\providecommand{\BIBentryALTinterwordstretchfactor}{4}
\providecommand{\BIBentryALTinterwordspacing}{\spaceskip=\fontdimen2\font plus
\BIBentryALTinterwordstretchfactor\fontdimen3\font minus
  \fontdimen4\font\relax}
\providecommand{\BIBforeignlanguage}[2]{{%
\expandafter\ifx\csname l@#1\endcsname\relax
\typeout{** WARNING: IEEEtran.bst: No hyphenation pattern has been}%
\typeout{** loaded for the language `#1'. Using the pattern for}%
\typeout{** the default language instead.}%
\else
\language=\csname l@#1\endcsname
\fi
#2}}
\providecommand{\BIBdecl}{\relax}
\BIBdecl

\bibitem{AV-TEST2023}
``{Malware},'' \url{https://www.av-test.org/en/statistics/malware/}, 2024.

\bibitem{TrendReports}
P.~P. Fyodor~Yarochkin, Zhengyu~Dong, ``{Lemon Group’s Cybercriminal
  Businesses Built on Preinfected Devices},''
  \url{https://www.trendmicro.com/en_us/research/23/e/lemon-group-cybercriminal-businesses-built-on-preinfected-devices.html},
  2023.

\bibitem{shi2020vahunt}
L.~Shi, J.~Ming, J.~Fu, G.~Peng, D.~Xu, K.~Gao, and X.~Pan, ``Vahunt: Warding
  off new repackaged android malware in app-virtualization's clothing,'' in
  \emph{Proceedings of the 2020 ACM SIGSAC Conference on Computer and
  Communications Security}, 2020, pp. 535--549.

\bibitem{arora2019permpair}
A.~Arora, S.~K. Peddoju, and M.~Conti, ``Permpair: Android malware detection
  using permission pairs,'' \emph{IEEE Transactions on Information Forensics
  and Security}, vol.~15, pp. 1968--1982, 2019.

\bibitem{alhanahnah2020dina}
M.~Alhanahnah, Q.~Yan, H.~Bagheri, H.~Zhou, Y.~Tsutano, W.~Srisa-An, and
  X.~Luo, ``Dina: Detecting hidden android inter-app communication in dynamic
  loaded code,'' \emph{IEEE Transactions on Information Forensics and
  Security}, vol.~15, pp. 2782--2797, 2020.

\bibitem{tsutano2019jitana}
Y.~Tsutano, S.~Bachala, W.~Srisa-an, G.~Rothermel, and J.~Dinh, ``Jitana: A
  modern hybrid program analysis framework for android platforms,''
  \emph{Journal of Computer Languages}, vol.~52, pp. 55--71, 2019.

\bibitem{zhang2020enhancing}
X.~Zhang, Y.~Zhang, M.~Zhong, D.~Ding, Y.~Cao, Y.~Zhang, M.~Zhang, and M.~Yang,
  ``Enhancing state-of-the-art classifiers with api semantics to detect evolved
  android malware,'' in \emph{Proceedings of the 2020 ACM SIGSAC conference on
  computer and communications security}, 2020, pp. 757--770.

\bibitem{samhi2022difuzer}
J.~Samhi, L.~Li, T.~F. Bissyand{\'e}, and J.~Klein, ``Difuzer: uncovering
  suspicious hidden sensitive operations in android apps,'' in \emph{2022
  IEEE/ACM 44th International Conference on Software Engineering (ICSE)}.\hskip
  1em plus 0.5em minus 0.4em\relax IEEE, 2022, pp. 723--735.

\bibitem{hou2017hindroid}
S.~Hou, Y.~Ye, Y.~Song, and M.~Abdulhayoglu, ``Hindroid: An intelligent android
  malware detection system based on structured heterogeneous information
  network,'' in \emph{Proceedings of the 23rd ACM SIGKDD international
  conference on knowledge discovery and data mining}, 2017, pp. 1507--1515.

\bibitem{onwuzurike2019mamadroid}
L.~Onwuzurike, E.~Mariconti, P.~Andriotis, E.~D. Cristofaro, G.~Ross, and
  G.~Stringhini, ``Mamadroid: Detecting android malware by building markov
  chains of behavioral models (extended version),'' \emph{ACM Transactions on
  Privacy and Security (TOPS)}, vol.~22, no.~2, pp. 1--34, 2019.

\bibitem{arp2014drebin}
D.~Arp, M.~Spreitzenbarth, M.~Hubner, H.~Gascon, K.~Rieck, and C.~Siemens,
  ``Drebin: Effective and explainable detection of android malware in your
  pocket.'' in \emph{Ndss}, vol.~14, 2014, pp. 23--26.

\bibitem{chen2020software}
J.~Chen, K.~Hu, Y.~Yu, Z.~Chen, Q.~Xuan, Y.~Liu, and V.~Filkov, ``Software
  visualization and deep transfer learning for effective software defect
  prediction,'' in \emph{Proceedings of the ACM/IEEE 42nd international
  conference on software engineering}, 2020, pp. 578--589.

\bibitem{sun2021android}
T.~Sun, N.~Daoudi, K.~Allix, and T.~F. Bissyand{\'e}, ``Android malware
  detection: looking beyond dalvik bytecode,'' in \emph{2021 36th IEEE/ACM
  International Conference on Automated Software Engineering Workshops
  (ASEW)}.\hskip 1em plus 0.5em minus 0.4em\relax IEEE, 2021, pp. 34--39.

\bibitem{he2023resnext+}
Y.~He, X.~Kang, Q.~Yan, and E.~Li, ``Resnext+: Attention mechanisms based on
  resnext for malware detection and classification,'' \emph{IEEE Transactions
  on Information Forensics and Security}, 2023.

\bibitem{he2022msdroid}
Y.~He, Y.~Liu, L.~Wu, Z.~Yang, K.~Ren, and Z.~Qin, ``Msdroid: Identifying
  malicious snippets for android malware detection,'' \emph{IEEE Transactions
  on Dependable and Secure Computing}, 2022.

\bibitem{liu2023enhancing}
Z.~Liu, L.~F. Zhang, and Y.~Tang, ``Enhancing malware detection for android
  apps: Detecting fine-granularity malicious components,'' in \emph{2023 38th
  IEEE/ACM International Conference on Automated Software Engineering
  (ASE)}.\hskip 1em plus 0.5em minus 0.4em\relax IEEE, 2023, pp. 1212--1224.

\bibitem{darem2021visualization}
A.~Darem, J.~Abawajy, A.~Makkar, A.~Alhashmi, and S.~Alanazi, ``Visualization
  and deep-learning-based malware variant detection using opcode-level
  features,'' \emph{Future Generation Computer Systems}, vol. 125, pp.
  314--323, 2021.

\bibitem{jeon2020malware}
S.~Jeon and J.~Moon, ``Malware-detection method with a convolutional recurrent
  neural network using opcode sequences,'' \emph{Information Sciences}, vol.
  535, pp. 1--15, 2020.

\bibitem{allix2016androzoo}
K.~Allix, T.~F. Bissyand{\'e}, J.~Klein, and Y.~Le~Traon, ``Androzoo:
  Collecting millions of android apps for the research community,'' in
  \emph{MSR'16}.\hskip 1em plus 0.5em minus 0.4em\relax IEEE, 2016, pp.
  468--471.

\bibitem{DBLP:conf/pldi/ArztRFBBKTOM14}
S.~Arzt, S.~Rasthofer, C.~Fritz, E.~Bodden, A.~Bartel, J.~Klein, Y.~Le~Traon,
  D.~Octeau, and P.~McDaniel, ``Flowdroid: Precise context, flow, field,
  object-sensitive and lifecycle-aware taint analysis for android apps,''
  vol.~49, no.~6.\hskip 1em plus 0.5em minus 0.4em\relax ACM New York, NY, USA,
  2014, pp. 259--269.

\bibitem{Androguard}
``Androguard,'' \url{https://github.com/androguard/androguard}, 2024.

\bibitem{APKTool}
``Apktool,'' \url{https://apktool.org/}, 2024.

\bibitem{kipf2016semi}
T.~N. Kipf and M.~Welling, ``Semi-supervised classification with graph
  convolutional networks,'' \emph{arXiv preprint arXiv:1609.02907}, 2016.

\bibitem{kim2014convolutional}
Y.~Kim, ``Convolutional neural networks for sentence classification,''
  \emph{arXiv preprint arXiv:1408.5882}, 2014.

\bibitem{yu2017multi}
Z.~Yu, J.~Yu, J.~Fan, and D.~Tao, ``Multi-modal factorized bilinear pooling
  with co-attention learning for visual question answering,'' in
  \emph{Proceedings of the IEEE international conference on computer vision},
  2017, pp. 1821--1830.

\bibitem{vaswani2017attention}
A.~Vaswani, N.~Shazeer, N.~Parmar, J.~Uszkoreit, L.~Jones, A.~N. Gomez,
  {\L}.~Kaiser, and I.~Polosukhin, ``Attention is all you need,''
  \emph{Advances in neural information processing systems}, vol.~30, 2017.

\bibitem{kim2018multimodal}
T.~Kim, B.~Kang, M.~Rho, S.~Sezer, and E.~G. Im, ``A multimodal deep learning
  method for android malware detection using various features,'' \emph{IEEE
  Transactions on Information Forensics and Security}, vol.~14, no.~3, pp.
  773--788, 2018.

\bibitem{qiu2022cyber}
J.~Qiu, Q.-L. Han, W.~Luo, L.~Pan, S.~Nepal, J.~Zhang, and Y.~Xiang, ``Cyber
  code intelligence for android malware detection,'' \emph{IEEE Transactions on
  Cybernetics}, vol.~53, no.~1, pp. 617--627, 2022.

\bibitem{au2012pscout}
K.~W.~Y. Au, Y.~F. Zhou, Z.~Huang, and D.~Lie, ``Pscout: analyzing the android
  permission specification,'' in \emph{Proceedings of the 2012 ACM conference
  on Computer and communications security}, 2012, pp. 217--228.

\bibitem{2017DroidNative}
S.~Alam, Z.~Qu, R.~Riley, Y.~Chen, and V.~Rastogi, ``Droidnative: Automating
  and optimizing detection of android native code malware variants,''
  \emph{Computers \& Security}, vol.~65, no. MAR., pp. 230--246, 2017.

\bibitem{liu2022convnet}
Z.~Liu, H.~Mao, C.-Y. Wu, C.~Feichtenhofer, T.~Darrell, and S.~Xie, ``A convnet
  for the 2020s,'' in \emph{Proceedings of the IEEE/CVF conference on computer
  vision and pattern recognition}, 2022, pp. 11\,976--11\,986.

\bibitem{wu2022vulcnn}
Y.~Wu, D.~Zou, S.~Dou, W.~Yang, D.~Xu, and H.~Jin, ``Vulcnn: An image-inspired
  scalable vulnerability detection system,'' in \emph{Proceedings of the 44th
  International Conference on Software Engineering}, 2022, pp. 2365--2376.

\bibitem{2020Measuring}
L.~Yang, S.~Zhu, J.~Shi, B.~Qin, and G.~Wang, ``Measuring and modeling the
  label dynamics of online anti-malware engines,'' in \emph{2020 USENIX
  Security Syposium}, 2020.

\bibitem{grosse2017adversarial}
K.~Grosse, N.~Papernot, P.~Manoharan, M.~Backes, and P.~McDaniel, ``Adversarial
  examples for malware detection,'' in \emph{Computer Security--ESORICS 2017:
  22nd European Symposium on Research in Computer Security, Oslo, Norway,
  September 11-15, 2017, Proceedings, Part II 22}.\hskip 1em plus 0.5em minus
  0.4em\relax Springer, 2017, pp. 62--79.

\bibitem{kang2016n}
B.~Kang, S.~Y. Yerima, K.~McLaughlin, and S.~Sezer, ``N-opcode analysis for
  android malware classification and categorization,'' in \emph{2016
  International conference on cyber security and protection of digital services
  (cyber security)}.\hskip 1em plus 0.5em minus 0.4em\relax IEEE, 2016, pp.
  1--7.

\bibitem{pendlebury2019tesseract}
F.~Pendlebury, F.~Pierazzi, R.~Jordaney, J.~Kinder, and L.~Cavallaro,
  ``$\{$TESSERACT$\}$: Eliminating experimental bias in malware classification
  across space and time,'' in \emph{28th USENIX security symposium (USENIX
  Security 19)}, 2019, pp. 729--746.

\bibitem{sebastian2016avclass}
M.~Sebasti{\'a}n, R.~Rivera, P.~Kotzias, and J.~Caballero, ``Avclass: A tool
  for massive malware labeling,'' in \emph{Research in Attacks, Intrusions, and
  Defenses: 19th International Symposium, RAID 2016, Paris, France, September
  19-21, 2016, Proceedings 19}.\hskip 1em plus 0.5em minus 0.4em\relax
  Springer, 2016, pp. 230--253.

\bibitem{ying2019gnnexplainer}
Z.~Ying, D.~Bourgeois, J.~You, M.~Zitnik, and J.~Leskovec, ``Gnnexplainer:
  Generating explanations for graph neural networks,'' \emph{Advances in neural
  information processing systems}, vol.~32, 2019.

\bibitem{selvaraju2017grad}
R.~R. Selvaraju, M.~Cogswell, A.~Das, R.~Vedantam, D.~Parikh, and D.~Batra,
  ``Grad-cam: Visual explanations from deep networks via gradient-based
  localization,'' in \emph{Proceedings of the IEEE international conference on
  computer vision}, 2017, pp. 618--626.

\bibitem{DBLP:conf/icse/OcteauLDJM15}
D.~Octeau, D.~Luchaup, M.~Dering, S.~Jha, and P.~McDaniel, ``Composite constant
  propagation: Application to android inter-component communication analysis,''
  in \emph{ICSE'15}, vol.~1.\hskip 1em plus 0.5em minus 0.4em\relax IEEE, 2015,
  pp. 77--88.

\bibitem{gao2023obfuscation}
C.~Gao, M.~Cai, S.~Yin, G.~Huang, H.~Li, W.~Yuan, and X.~Luo,
  ``Obfuscation-resilient android malware analysis based on complementary
  features,'' \emph{IEEE Transactions on Information Forensics and Security},
  2023.

\bibitem{gao2024comprehensive}
C.~Gao, G.~Huang, H.~Li, B.~Wu, Y.~Wu, and W.~Yuan, ``A comprehensive study of
  learning-based android malware detectors under challenging environments,'' in
  \emph{Proceedings of the 46th IEEE/ACM International Conference on Software
  Engineering}, 2024, pp. 1--13.

\bibitem{hou2021disentangled}
S.~Hou, Y.~Fan, M.~Ju, Y.~Ye, W.~Wan, K.~Wang, Y.~Mei, Q.~Xiong, and F.~Shao,
  ``Disentangled representation learning in heterogeneous information network
  for large-scale android malware detection in the covid-19 era and beyond,''
  in \emph{Proceedings of the AAAI Conference on Artificial Intelligence},
  vol.~35, no.~9, 2021, pp. 7754--7761.

\bibitem{pauck2018android}
F.~Pauck, E.~Bodden, and H.~Wehrheim, ``Do android taint analysis tools keep
  their promises?'' in \emph{Proceedings of the 2018 26th ACM joint meeting on
  european software engineering conference and symposium on the foundations of
  software engineering}, 2018, pp. 331--341.

\bibitem{backes2016demystifying}
M.~Backes, S.~Bugiel, E.~Derr, P.~McDaniel, D.~Octeau, and S.~Weisgerber, ``On
  demystifying the android application framework:$\{$Re-Visiting$\}$ android
  permission specification analysis,'' in \emph{25th USENIX security symposium
  (USENIX security 16)}, 2016, pp. 1101--1118.

\bibitem{hoffmann2013slicing}
J.~Hoffmann, M.~Ussath, T.~Holz, and M.~Spreitzenbarth, ``Slicing droids:
  program slicing for smali code,'' in \emph{Proceedings of the 28th Annual ACM
  Symposium on Applied Computing}, 2013, pp. 1844--1851.

\bibitem{wu2021homdroid}
Y.~Wu, D.~Zou, W.~Yang, X.~Li, and H.~Jin, ``Homdroid: detecting android covert
  malware by social-network homophily analysis,'' in \emph{Proceedings of the
  30th acm sigsoft international symposium on software testing and analysis},
  2021, pp. 216--229.

\bibitem{narayanan2018multi}
A.~Narayanan, M.~Chandramohan, L.~Chen, and Y.~Liu, ``A multi-view
  context-aware approach to android malware detection and malicious code
  localization,'' \emph{Empirical Software Engineering}, vol.~23, no.~3, pp.
  1222--1274, 2018.

\bibitem{narayanan2018apk2vec}
A.~Narayanan, C.~Soh, L.~Chen, Y.~Liu, and L.~Wang, ``apk2vec: Semi-supervised
  multi-view representation learning for profiling android applications,'' in
  \emph{2018 IEEE International Conference on Data Mining (ICDM)}.\hskip 1em
  plus 0.5em minus 0.4em\relax IEEE, 2018, pp. 357--366.

\bibitem{niu2023gcdroid}
W.~Niu, Y.~Wang, X.~Liu, R.~Yan, X.~Li, and X.~Zhang, ``Gcdroid: Android
  malware detection based on graph compression with reachability relationship
  extraction for iot devices,'' \emph{IEEE Internet of Things Journal}, 2023.

\end{thebibliography}
